\documentclass[aps, pre, onecolumn, nofootinbib, notitlepage, groupedaddress, longbibliography, superscriptaddress]{revtex4-1}
\usepackage{amsmath}
\usepackage{amssymb}
\usepackage{amsfonts}
\usepackage{graphicx}
\usepackage{hyperref}
\usepackage{xcolor}
\usepackage{ulem}

\usepackage{csvsimple}
\usepackage{longtable}

\hypersetup{
    colorlinks,
    linkcolor={red!50!black},
    citecolor={blue!50!black},
    urlcolor={blue!80!black}
}

\newcommand{\grad}{\ensuremath{\nabla}}

\newcommand\apjl{{The Astrophysical Journal Letters}}
\newcommand\apjs{{The Astrophysical Journal Supplement}}
\newcommand\ssr{{Space Science Reviews}}
\newcommand\aap{{Astronomy and Astrophysics}}
\newcommand\grl{{Geophysics Research Letters}}
\newcommand\mnras{{Monthly Notices of the Royal Astronomical Society}}

\newcommand{\Heat}{\ensuremath{\mathcal{H}}}
\newcommand{\angles}[1]{\ensuremath{\left\langle#1\right\rangle}}
\renewcommand{\vec}[1]{\ensuremath{\mathbf{#1}}}

\usepackage{xcolor}
\usepackage{ulem}

\newcommand{\comment}[1]{}

\makeatletter
\def\input@path{{./sections/}}
\makeatother
\graphicspath{{./}{figures/}}

\begin{document}

\title{Internally heated and fully compressible convection: flow morphology and scaling laws}

\author{Whitney T. Powers}
\affiliation{Department of Astrophysical and Planetary Sciences \\
University of Colorado Boulder}
\author{Evan H. Anders}
\affiliation{Center for Interdisciplinary Exploration and Research in Astrophysics (CIERA), Northwestern University, 1800 Sherman Ave, Evanston, IL 60201,USA}
\author{Benjamin P. Brown}
\affiliation{Department of Astrophysical and Planetary Sciences \\
University of Colorado Boulder}
\normalem
\begin{abstract}
In stars and planets natural processes heat convective flows in the bulk of a convective region rather than at hard boundaries. By characterizing how convective dynamics are determined by the strength of an internal heating source we can gain insight into the processes driving astrophysical convection. Internally heated convection has been studied extensively in incompressible fluids, but the effects of stratification and compressibility have not been examined in detail. In this work, we study fully compressible convection driven by a spatially uniform heating source in 2D and 3D Cartesian, hydrodynamic simulations. We use a fixed temperature upper boundary condition which results in a system that is internally heated in the bulk and cooled at the top. We find that the flow speed, as measured by the Mach number, and turbulence, as measured by the Reynolds number, can be independently controlled by separately varying the characteristic temperature gradient from internal heating and the diffusivities. 2D simulations at a fixed Mach number (flow speed) demonstrate consistent power at low wavenumber as diffusivities are decreased. We observe convection where the velocity distribution is skewed towards cold, fast downflows, and that the flow speed is related to the length scale and entropy gradient of the upper boundary where the downflows are driven. We additionally find a heat transport scaling law which is consistent with prior incompressible work.

\end{abstract}

\maketitle

\section{Introduction}
\label{sec:intro}

Turbulent convective flows are critical to many astrophysical phenomena. In planets and brown dwarfs, convection drives observable chemical disequilibrium \citep{Li2017,Li2020JupWater,Bordwell} and cloud formation \citep{Vallis2019RainyBenard, Lefevre2022Cloud}. In stars, convection generates differential rotation \citep{Thompson1996SolarDifRot, Hotta2021Nature} and magnetic dynamos \citep{Brown2011dynamo, Charbonneau2020review}. While convection is common, it remains poorly understood, as is indicated by the solar convective conundrum \citep{Rast2020ConvectiveConundrum, Vasil2021ConvectiveConundrum}, where reported helioseismic observations of convective flows differ by orders of magnitude from one another and from theory and simulations \citep{Hanasoge2012, Hanasoge2015, Greer2015, proxauf2021thesis_conumdrum}. This disagreement calls for a re-examination of the fundamental processes driving solar convection. In order to understand emergent convectively-driven phenomena in stars and planets, we must develop robust models of the convection driving them.

Astrophysical convection is often driven by internal heating. In stars, convection is caused by changes in opacity and nuclear fusion, both of which enter the equations of motion as spatially distributed heating processes \citep[e.g.,][]{Jermyn2022ConvectionAtlas}. In gas giant planets, convection is driven by changes in radiative conductivity and the latent heat of formation \citep{Showman2020review}. The evolution time of stars is much longer than the dynamical times for convection, and so a non-uniform conductivity can be treated like a stationary-in-time source of internal heating. Simulations which attempt to model astrophysical systems include many physical processes, one of which is internal heating \citep{Meakin_Arnet2007, Woodward2015, Herwig2023_falkI, Thompson2023_falkII, Blouin2023_falkIII, Blouin2023_falkIV}. While many complex simulations drive convection with internal heating, these studies usually examine how the heating strength affects processes including boundary mixing and waves in adjacent stable regions, and there has been less work studying how heating strength affects the fundamental flow measures in fully compressible, stratified fluids \citep{Cristini2019HeatedBoundaryMixing, baraffe_etal_2021, LeSaux2022_baraffe21II, Baraffe2023_overshoot}. It is important to understand how the input parameters of internally heated convection map to important measures such as Mach number number so that we can produce suites of simulations where Mach number is held constant at a known value for a particular system.

In many experiments and numerical simulations, convection is driven at hot or cold boundaries, and the resulting boundary layers throttle the heat transport \citep{Grossmann_Lohse_2000, Ahlers2009RvMP...81..503A}. Some recent simulations and experiments of incompressible fluids instead drive convection using internal heating in an effort to reduce boundary effects. Numerical simulations from \citet{goluskin_spiegel_2012} and \citet{Goluskin_vanderPoel_2016JFM}  measured heat transport scaling laws in internally heated Boussinesq convection, and these findings were generalized to find mathematically robust limits to heat transport scalings in \citet{Goluskin2015} and diffusion-based bounds in \citet{Wang2021IHC}. \citet{Miquel2019BeyondUltimate} found that internally heated convection with a non-uniform heat source can, surprisingly, produce heat transport scaling in excess of the limit set by mixing length theory. Several laboratory experiments have studied internally heated convection using a dye to absorb light thereby heating the fluid and have found that heat transport scaling depends on the concentration of the dye and thus the length scale over which the fluid is heated \citep{Lepot2018IntHeatExp,Bouillaut2019IHCExperiment}. Earlier internal heating experiments used electrical currents for bulk heating, and incorporated a cooled upper boundary \citep{kulacki_goldstein_1972IntHeatExp}. The approach of using internal heating and cooling has also found its way into some studies of astrophysical convection driven through heating and cooling layers since it produces diffusion free scalings, which are relevant in astrophysical settings where bulk heating processes drive convective flows without hard boundaries \citep{barker_etal_2014, Currie2020MisalignedRotation}. 


In this work we study fully compressible convection which is internally heated in the bulk and cooled at the top boundary. This is accomplished using a constant and uniform source term of internal energy acting upon an adiabatically stratified, ideal gas atmosphere. We enforce a fixed temperature upper boundary, giving rise to boundary cooling at the top of the domain. This setup has been studied in the case of Boussinesq convection in the uniform heating case of \citet{kazemi_etal_2021}. This case is also analogous to the IH3 configuration described in \citet{Goluskin2015}. This setup resembles convection near the photosphere of lower main sequence and red giant branch stars, where rapid photospheric cooling can be approximated as boundary cooling. 

We characterize internally heated fully compressible convection, and in particular how quantities like the Mach number (Ma) and Reynolds number (Re) vary with our chosen input parameters. Mach number may be known for a particular system, and in this case it is important to have a suite of simulations with this particular Mach number and different Reynolds numbers. Mach number is measure of how important compressibility is to the dynamics, and thus when alternative equation sets such as the anelastic or Boussinesq can be used. In this work we determine the input parameters that allow us to change Mach number and Reynolds number independently.

We demonstrate that changing the magnitude of the heating term affects the Mach number, and thus the importance of compressibility. With this setup, flow speed as measured by the Mach number, and turbulence as measured by the Reynolds number, can be controlled independently in 2D simulations and are nearly independent in 3D. We observe convective flows dominated by cool downflows produced at the upper boundary layer. At fixed heating magnitude, we find that heat transport and the Reynolds number follow expected scaling laws from Boussinesq convection. With independent control of Mach number and Reynolds number we can produce a series of simulations with consistent behavior in convective power spectra at low wavenumber.

\section{Methods}
\label{sec:methods}
We study the evolution of an ideal gas according to the fully compressible Navier-Stokes equations. We follow previous works \citep[e.g.,][]{lecoanet_etal_2014, AndersBrown2017} and use a temperature and log-density formulation of the equations. We modify the equations by adding a  heating term, $Q$, to the internal energy equation. These fully compressible equations are:
    \begin{gather}
        \frac{\partial \ln \rho}{\partial t}  + \textbf{u}\cdot \grad \ln\rho  + \grad \cdot \textbf{u} =0,    
        \label{eqn:continuity}
\\
        \frac{\partial \textbf{u}}{\partial t} + \textbf{u}\cdot \grad \textbf{u} = - 
        \frac{R}{m}\left[ \grad T + T \grad \ln \rho \right] +\vec{g}+ \frac{\mu}{\rho} \grad \cdot \sigma,
        \label{eqn:momentum}
\\
            \frac{\partial T}{\partial t} + \textbf{u}\cdot \grad T + (\gamma-1)T\grad\cdot \textbf{u} = \frac{1}{\rho c_v}\mu\Phi + \frac{\kappa}{\rho c_v} \grad^2 T + \frac{1}{\rho c_v}Q,
            \label{eqn:temperature}
    \end{gather}
where $\rho$ is the density, $\mathbf{u} = u \hat{x} + v \hat{y} + w\hat{z}$ is the velocity vector, $T$ is temperature,  $\mu$ is the dynamic viscosity, $\kappa$ is the thermal diffusion, and $Q$ is an internal heating source term. We take $\mu$, $\kappa$, and $Q$ to be constant in time and spatially uniform, which is assumed in the expressions of equations \ref{eqn:momentum}-\ref{eqn:temperature}. 
We use an ideal gas equation of state for pressure $P = \frac{R} {m} \rho T$, where $R$ is the gas constant and $m$ is the mean molecular weight. $c_V$ and $c_P$ are the specific heat at constant volume and pressure respectively, and their ratio is the adiabatic exponent $\gamma=\frac{c_P}{c_V}$.
The viscous stress tensor $\sigma$ and viscous heating term $\Phi$ are respectively defined as
\begin{gather}
     \sigma_{ij} = 2\left[E_{ij}-\frac{1}{3}(\grad\cdot \mathbf{u})  I_{ij}\right],
     \label{eqn:stress_tensor}
\\
    \Phi = 2 \left(E_{ij}E_{ij} - \frac{1}{3}(\grad \cdot \mathbf{u})^2 \right),
    \label{eqn:visc_heating}
\end{gather}
where $I_{ij}$ is the identity tensor and
\begin{equation}
    E_{ij} = \frac{1}{2} \left[\grad \mathbf{u} +(\grad \mathbf{u})^T\right]
    \label{eqn:strain_rate}
\end{equation}
is the rate-of-strain tensor.
We decompose our thermodynamics into an adiabatic and hydrostatically equilibrated portion (which we denote with subscripts 0) and fluctuations around that background (which we denote with subscripts 1). We decompose our fully non-linear thermodynamics about that background state, $\ln \rho = \ln \rho_0 + \ln \rho_1$ and $T = T_0+T_1$. The temperature gradient of the hydrostatic background is set to the adiabatic temperature gradient, $\grad T_0 = \grad T_\text{ad} = \mathbf{g}/c_P$ where $\mathbf{g} =-g \hat{z} $ is the gravitational acceleration vector which is taken to be constant and uniform. The domain spans $z \in [0, L_z]$ and is a horizontally periodic box with aspect ratio $\Gamma = L_x/L_z = 4$. For 3D simulations we set $L_x=L_y$. At the top and bottom boundary we enforce no-slip boundary conditions,
\begin{equation}
    \vec{u}(z=0) = \vec{u}(z=L_z) = 0.
    \label{eqn:velocity_BC}
\end{equation}
with mixed thermal boundary conditions, where we fix the temperature gradient at the bottom and fix the temperature at the top of the domain,
\begin{equation}
    \frac{\partial T_1}{\partial z}(z=0) = T_1(z=L_z) = 0.
    \label{eqn:thermal_BC}
\end{equation}

We nondimensionalize our equations using values of $\rho$ and $T$ taken at top of the domain to specify our reference temperature, $T_{c}$, and density $\rho_{c}$ such that $T_0(z=L_z) = 1$ and $\ln\rho_0(z=L_z)=0$. We choose a nondimensional time scale by setting the ideal gas constant $R/m=1$. This choice ensures that the isothermal sound speed, $c_s^2 = \partial P / \partial \rho = R/m T = 1$ when evaluated at the top of the domain. We choose our characteristic length scale to be $L_c = T_c/{\grad T_\mathrm{ad}}$. We specify the depth of our domain so that the adiabatic reference state spans $N_\rho = 3$ density scale heights according to the formula:
\begin{equation}
 L_z = e^{(\gamma - 1)N_\rho} -1.   
\end{equation}
We set the ratio of viscous to thermal diffusivities with the Prandtl number $\text{Pr}  = \mu c_P /\kappa = 1$. 

\begin{figure}
    \centering
    \includegraphics[width=\textwidth]{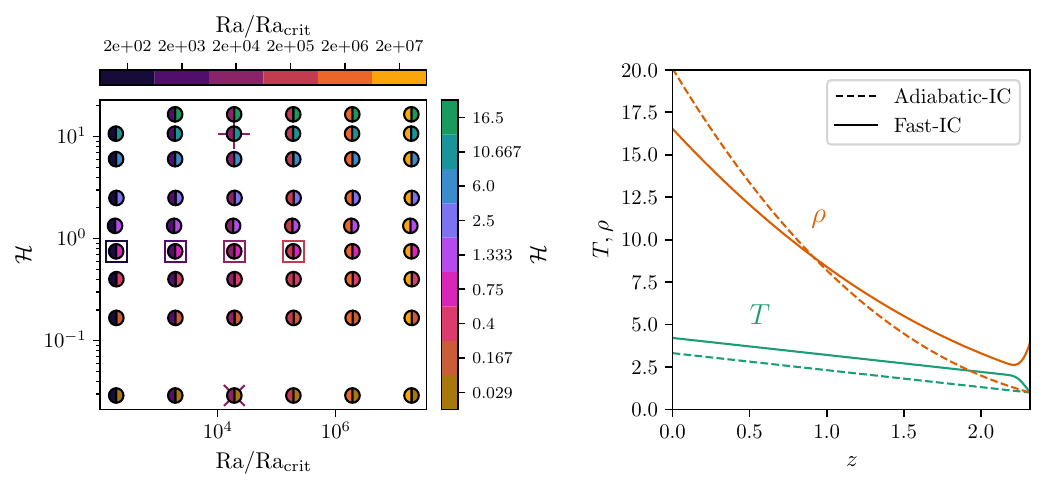}
    \caption{ (Left panel) Symbols indicate the simulations conducted in this work in $\Heat$ - $\mathrm{Ra}/\mathrm{Ra}_\mathrm{crit}$ parameter space. 2D simulations are indicated with a solid circle. 3D simulations are indicated with the square outline, $\times$, and $+$ symbols. The marker color indicates the value of Ra and $\Heat$ for each point. The left side of each 2D point indicates the value of Ra (upper color bar) and the right side indicates the value of $\Heat$ (right color bar). 3D points are colored by the value of Ra. While these colors are redundant with the position on the plot, these colors are used consistently throughout the rest of the figures to refer to $\mathrm{Ra}/\mathrm{Ra}_\mathrm{crit}$ and $\Heat$. (Right panel) Initial conditions for $T$ and $\rho$ where dashed lines correspond to an adiabatic polytrope stratification and solid lines show the ``Fast-IC" initial conditions, where we approximate the temperature structure of the final state and solve a boundary value problem for the density profile which satisfies hydrostatic equilibrium and mass conservation, see Appendix \ref{sec:apdx_IC}. The ``Fast-IC" profile shown is for $\mathrm{Ra}/\mathrm{Ra}_\mathrm{crit}=2\times10^5$ and $\Heat=10.667$.}
    \label{fig:roadmap_IC}
\end{figure}

To control the simulations we specify the strength of the source term in the energy equation with a nondimensional heating strength parameter, $\Heat$, such that
\begin{equation}
    Q=\frac{\kappa}{L_z} \Heat,
    \label{eqn:Q_defn}
\end{equation}
where $\Heat$ is constant. $\Heat$ measures a characteristic temperature gradient scale from the heat source, $\Heat = \nabla T_1 (z=L_z)/\nabla T_ad$. Since we are considering a constant and uniform heat source, the flux in excess of that carried by the adiabat linearly increases with height and the slope of the flux is determined by \Heat.

By separating $Q$ into a part proportional to $\kappa$ and a part proportional to $\Heat$ we ensure that the hydrostatic thermal equilibrium 
\begin{equation}
    \nabla^2 T = -\frac{\Heat}{L_z}
    \label{eqn:static_T}
\end{equation}
remains constant as we change the degree of turbulence via changes to the diffusivities. By integration we can produce a constant temperature scale for this static solution
\begin{equation}
    \Delta T_\mathrm{static} = 2\frac{L_z^2 Q}{\kappa} = 2L_z \Heat.
\end{equation}
The quantity $\Delta T_\mathrm{static}$ is analagous to the quantity $\Delta$ in equation 1.5 of \citet{Goluskin2015}. Using $\Heat$ as an input parameter removes the scaling factor of kappa from the heating function, which ensures that the Rayleigh number only depends on diffusivities when $\Heat$ is fixed.

 We specify a Rayleigh number at the top of the domain to control the degree of turbulence in a simulation. The Rayleigh number, Ra, takes the form 
\begin{equation}
    \mathrm{Ra} = \frac{g L_z^3 \rho_c^2 c_P (\Delta T_\mathrm{static}/2T_{c})}{\mu \kappa} = \frac{g L_z^4 \rho_c^2 c_{P}}{\mu \kappa} \left( \frac{\Heat}{T_{c}}\right).
\end{equation}
$\Heat$ is a fundamental input parameter, and the Ra here is directly comparable to the one used to study Boussinesq, internally heated convection in  \citet{goluskin_spiegel_2012}, \citet{Goluskin2015}, and \citet{kazemi_etal_2021}. We perform a linear stability analysis using eigentools \citep{eigentools} to determine the value of $\mathrm{Ra}_{\rm{crit}}$; see appendix \ref{app:ra_crit} for details on how we perform this analysis. We find that the critical Ra is near constant at $\mathrm{Ra}_\mathrm{crit}\approx50$ for values of
 $\Heat<1$. At higher values of $\Heat$ the critical Rayleigh number increases with $\Heat$. 
 
There are 8 nondimensional numbers which control this experiment, Rayleigh number, $\Heat$, Prandtl number, $\gamma$, $N_\rho$, 2 aspect ratios (for x and y), and $\nabla T (z=0)/\nabla T_\mathrm{ad}$ which we have set to 1 implicitly with our setup. We derive the number of nondimensional parameters in a similar manner to \citet{Graham1975JFM}, however we have two additional nondimesional parameters from the addition of a 3rd dimension and our internal heat source. The non-dimensionalized equations are
    \begin{gather}
        \frac{\partial \ln \rho'}{\partial t}  + \textbf{u}'\cdot \grad \ln\rho'  + \grad \cdot \textbf{u}' =0,    
        \label{eqn:continuity_nd}
\\
        \frac{\partial \textbf{u}'}{\partial t} + \textbf{u}'\cdot \grad \textbf{u}' = - \left[ \grad T' + T' \grad \ln \rho' \right] -\frac{\gamma}{\gamma-1}\hat{z} + L_z^2 \sqrt{\frac{\mathcal{H}\mathrm{Pr}}{\mathrm{Ra}}}\frac{1}{\rho'} \grad \cdot \sigma',
        \label{eqn:momentum_nd}
\\
            \frac{\partial T'}{\partial t} + \textbf{u}'\cdot \grad T' + (\gamma-1)T'\grad\cdot \textbf{u}' = (\gamma-1)L_z^2\sqrt{\frac{\mathcal{H}\mathrm{Pr}}{\mathrm{Ra}}}\frac{1}{\rho' }\Phi' + \gamma L_z^2\sqrt{\frac{\mathcal{H}}{\mathrm{Ra}\mathrm{Pr}}}\frac{1}{\rho'} \left[\grad^2 T' + \frac{\mathcal{H}}{L_z}\right].
            \label{eqn:temperature_nd}
    \end{gather}

For our numerical experiment we vary Ra and $\Heat$ and hold all other nondimensional inputs constant. With this choice of input parameters, varying Ra at constant $\Heat$ changes the value of $\kappa$ and $\mu$, while changing $\Heat$ at constant Ra changes the magnitude of the source term while also changing $\kappa$ and $\mu$ to ensure that the ratio of the buoyancy timescale to the diffusive timescale remains constant. By using Ra and $\Heat$ as our fundamental input parameters, we are able to largely separate the posterior Reynolds and Mach number of our simulations.  We run 2D simulations across 9 values of $\Heat$ and 6 values of $\text{Ra} / \text{Ra}_\text{crit}$. These 53 simulations are shown in parameter space in the upper left panel of Fig. \ref{fig:roadmap_IC}. We run select 3D simulations to validate the behavior of the 2D simulations; these simulations are denoted as boxes, $\times$, and $+$ to indicate different values of $\Heat$.

We decompose equations \ref{eqn:continuity_nd} - \ref{eqn:temperature_nd} into linear and non-linear terms using the adiabatic background state and  then use an implicit-explicit timestepper to implicitly timestep the linear terms and explicitly timestep the nonlinear terms. Our timestep is therefore determined by the CFL constraint of the nonlinear advective term rather than the linear sound waves or diffusion terms, enabling us to simulate low Mach flows efficiently. We timestep the decomposed version of equations \ref{eqn:continuity_nd}-\ref{eqn:temperature_nd} with the Dedalus v2 \footnote{Commit with short-SHA 1339061} pseudospectral solver \citep{Burns2020} at resolutions up to $2048\times1024$ spectral coefficients for 2D simulations and up to $512\times512\times256$ for 3D simulations. We use an implicit-explicit, four-stage, third-order Runge-Kutta RK443 timestepper \citep{ASCHER1997} with a CFL safety factor of $0.75$ for 2D simulations and the two-step semi-implicit backwards differentiation SBDF2 timestepper \citep{Wang2008} with a CFL safety factor of $0.3$ for 3D simulations. All fields are represented as spectral expansions of $N_z$ Chebyshev coefficients in the vertical ($z$) basis, and $N_x, N_y$ Fourier coefficients in the horizontal ($x$ and $y$) bases thus making our domain horizontally periodic. To avoid aliasing errors, we use the 3/2-dealiasing rule in all directions. We initialize the simulations with random noise temperature perturbations with magnitude $10^{-3}Q^{1/3}\sin(\pi z / L_z)$ added to the initial temperature profile. We define a heating timescale from mixing length theory $t_h = Q^{-1/3}$ to set the output cadence for the simulations. The code used in this paper can be found at \url{https://github.com/whitney-powers/internally_heated_fc_convection}. 

Simulations whose initial temperature profiles are the adiabatic state $T_0$ experience a long thermal relaxation time while a superadiabatic thermal boundary layer develops at the upper boundary. To reduce the time spent bringing simulations into thermal equilibrium, we construct an initial stratification characterized by a temperature profile which is adiabatic in the bulk but has a superadiabatic upper boundary layer. We will refer to these initial conditions as ``fast-IC". The width of the boundary layer is predicted from a Nusselt number scaling law extrapolated from low-Ra, low resolution simulations which used an adiabatic polytrope for their initial stratification. We solve a nonlinear boundary value problem to find the density stratification for a specified temperature profile which satisfies hydrostatic equilibrium while containing the same total mass as the adiabatic initial conditions. Example stratifications for adiabatic and fast initial conditions are shown in the right panel of Fig. \ref{fig:roadmap_IC}. 

The ``Fast-IC" approach offers significant improvement in the run time of 2D simulations. We discuss the convergence and performance properties of ``Fast-IC'' in more detail in section \ref{subsec:scalings}. We use fast-IC for 2D cases with $\mathrm{Ra}/\mathrm{Ra}_\mathrm{crit} > 2 \times 10^5$ and for all 3D cases except the $\mathcal{H}=10.667$. We calibrated the fast-IC by measuring and extrapolating a Nu (Nusselt number) vs Ra power law (Eqn. \ref{eqn:nondim_outputs})from 2D simulations with $\mathrm{Ra}/\mathrm{Ra}_\mathrm{crit} \leq 2 \times 10^5$. For each value of  $\Heat$ we fit a scaling law for $\mathrm{Nu}$ of form $\mathrm{Nu}= A* \mathrm{Ra}^\alpha$, where $\alpha\approx1/5$ and the prefactor $A \approx 1$. The exact values of $A$ and $\alpha$ are listed in Appendix \ref{app:initial_conditions}. This extrapolation works well for 2D, but for 3D runs the 2D calibration produces initial conditions with too large of a thermal boundary layer implying a subtly different Nu vs. Ra law in 2D and 3D. This discrepancy is especially pronounced at high $\Heat$ where these initial conditions produce a highly unstable region which leads to non-physical negative temperature perturbations which cause a timestepping instability. For the 3D, $\mathrm{Ra}=2\times 10^4 \mathrm{Ra}_\mathrm{crit}$, $\Heat=10.667$ case we reverted to the standard adiabatic polytrope for the initial conditions. Nu vs.~Ra in the fast-IC approach should be properly calibrated from 3D results to improve performance. For the details of generating ``fast-IC" initial conditions see Appendix \ref{app:initial_conditions}.

\section{Results}
\label{sec:results}
\subsection{Dynamics}
\label{subsec:dynamics}
\begin{figure}
    \centering
    \includegraphics[ width =\textwidth]{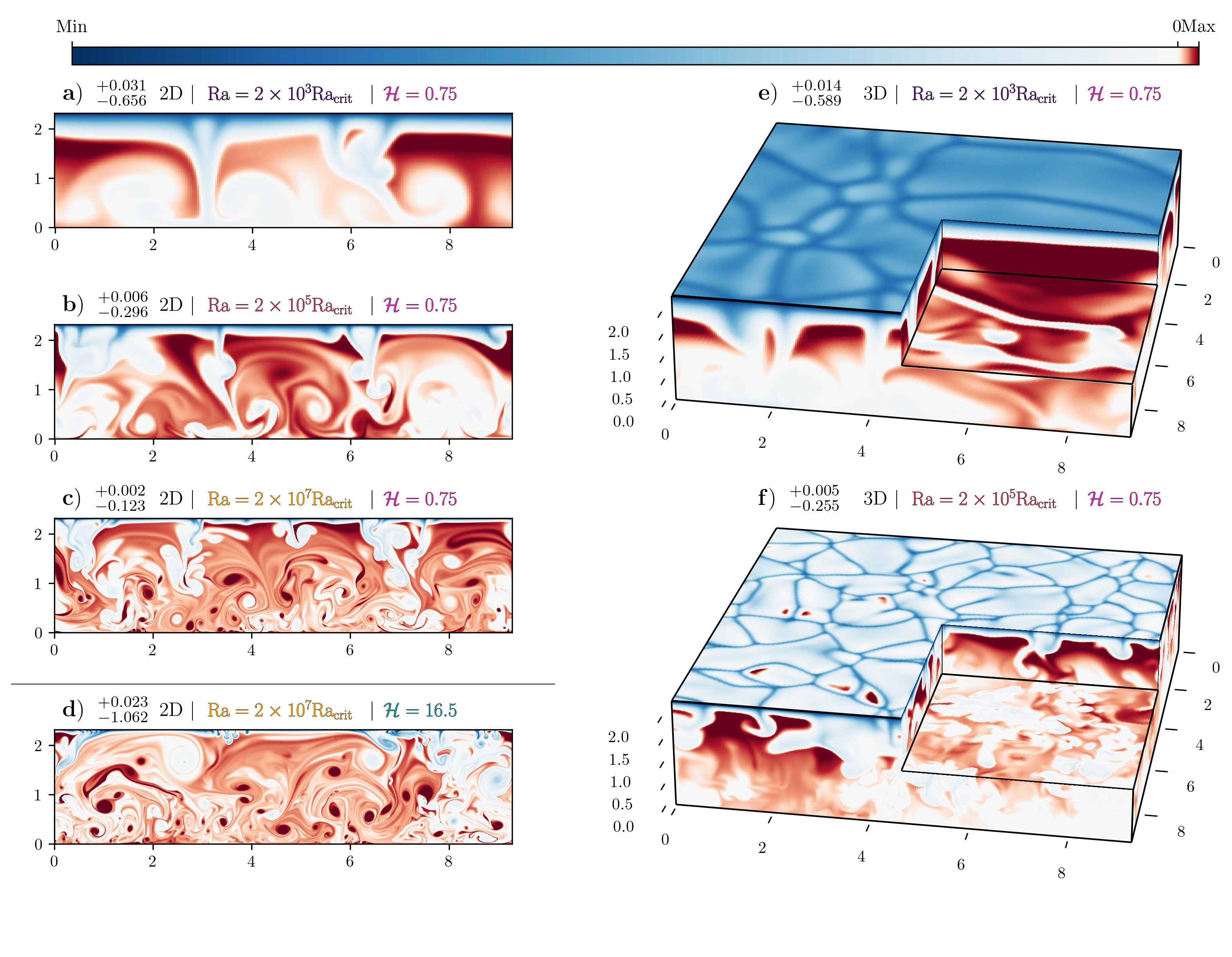}
    \caption{ We show entropy fluctuations with the adiabatic entropy subtracted. For each panel we set the colorbar limits at the 3rd and 97th percentile values of entropy from that simulation. These minimum and maximum values are displayed above each panel. Note the substantial imbalance between the extent of the minimum and maximum values. We show 2D dynamics in the left column, and 3D dynamics in the right column. In panels a), b), and c) we show 2D dynamics at $\Heat=0.75$ and with $\text{Ra}/\text{Ra}_\text{crit} = 2\times 10^3,  2\times 10^5,$ and $2\times 10^7$, respectively. As Rayleigh number increases, the flows become more turbulent. In panel d) we show a simulation with the same Rayleigh number as panel c), but with stronger internal heating at $\Heat=16.5$. In panel e) we display the 3D simulation with the same parameters as panel a). and in panel f) we display the 3D simulation with the same parameters as panel b). An animation of this figure can be found at \url{https://vimeo.com/793788958}}
    
   
    \label{fig:dynamics}
\end{figure}

    In Fig. \ref{fig:dynamics} we visualize the entropy fluctuations about the adiabatic entropy in four 2D and two 3D simulations. Since entropy in the interior is constant (see Fig. \ref{fig:profiles}) we estimate this adiabatic entropy by taking a time averaged value at the mid z-plane.  We observe a cold (blue) upper boundary layer where heat is transported by conduction. In both 2D and 3D systems the dynamics are dominated by convective downflow lanes in blue while convective upflows in red are weaker in magnitude. We use an asymmetric colorbar so that features can be resolved, but note that features in red are at much lower magnitude than features in blue. We do not observe heated upflow plumes at the lower boundary, and the boundary itself is not heated as the superadiabatic flux is zero. Flows at the upper boundary appear similar to standard boundary driven flows with cold plumes launching from a cooled layer. The behavior of the hot and cold plumes is indicative of the internal heating. Downflows cool at the upper boundary and as they fall they are heated, becoming buoyantly neutral (white), and hot upflows gradually heat as they rise, becoming a more intense red. We observe flow morphologies which are similarly dominated by cold downflows in 2D and 3D simulations. Our high Rayleigh number 2D simulations have characteristic features of 2D turbulence, such as vortices in the hot upflow plumes and the formation of large flywheel structures, albeit these flywheels are weaker than is typically observed in boundary driven convection \citep{AndersBrown2017}. We observe multiple downflow plumes in highly turbulent states in contrast to the single plume flywheel structure previously observed in boundary driven convection.
    
    In Fig. \ref{fig:dynamics} a-c, we show how dynamics change with increasing supercriticality in 2D simulations at low  $\mathcal{H}$. The simulation in Fig. \ref{fig:dynamics} a is  laminar. As the Rayleigh number increases in Fig. \ref{fig:dynamics} b and c we see more complex turbulent dynamics emerge. These include long-lived wrapped vortices, where hot fluid is encircled by cold fluid indicating high P\'{e}clet number flows. 

    In Fig. \ref{fig:dynamics} d we show turbulent 2D dynamics with a higher value of $\Heat=16.5$. The dynamics appear similarly turbulent between Fig. \ref{fig:dynamics} c and d;  both have wrapped vorticies and similar levels of small scale detail. This indicates that $\Heat$ has little effect on the Reynolds number of 2D flows. Where these two simulations differ is in the magnitude of fluctuations. The fluctuations in panel d are an order of magnitude larger than in panel c as indicated by the colorbar limits (see the $\pm$ values left of ``2D''). So varying Ra and $\Heat$ control mostly independent properties of the dynamics. 
    
    In the right column we show the effect of increasing Rayleigh number in 3D. The simulation in Fig. \ref{fig:dynamics} e shares the same values of Ra and $\Heat$ as Fig. \ref{fig:dynamics} a, and Fig. \ref{fig:dynamics} f shares the same values as Fig. \ref{fig:dynamics} b. As 3D flows become more turbulent we see more convection cells at the upper boundary and a transition from coherent cell structure at the midplane to flows dominated by discrete downflow plumes distributed randomly across the midplane. The behavior of the cool downflows in Fig. \ref{fig:dynamics} f are consistent with weak upflows, as indenpendent downflow plumes are not pushed into coherent downflow lanes. 

\begin{figure}
    \centering
    \includegraphics{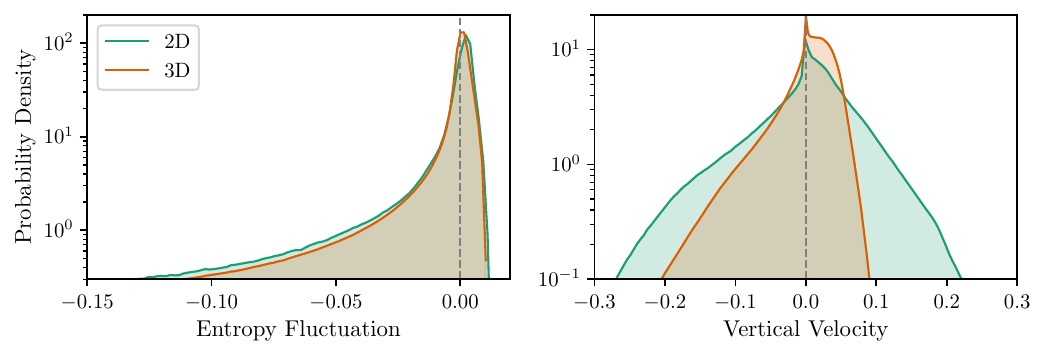}
    \caption{Time-averaged probability distribution functions (PDFs) of entropy fluctuations with the adiabatic entropy subtracted (left panel) and vertical velocity (right panel) are shown for 2D (green line) and 3D (orange line) simulations. Both simulations use  $\text{Ra}/\text{Ra}_\text{crit} = 2\times 10^5$ and $\Heat=0.75$. The PDFs are averaged over 12 heating timescales. The entropy PDF is skewed towards negative values and is nearly identical between the 2D and 3D case. 
    In contrast, the velocity PDFs exhibit different behavior in 2D and 3D. The 3D case has a narrower velocity distribution with more pronounced skew towards downward flows. }
    \label{fig:pdfs}
\end{figure}
The asymmetry between upflows and downflows can be observed more quantitatively through probability distribution functions (PDFs) of the entropy fluctuations and vertical velocity as shown in Fig. \ref{fig:pdfs}. We plot PDFs of a 2D (green line) and 3D (orange line) simulation at $\text{Ra}/\text{Ra}_\text{crit} = 2\times 10^5$ and $\Heat=0.75$ which correlates to panels b and f of Fig. \ref{fig:dynamics}. The PDF of entropy fluctuations about the adiabatic value(left panel) are nearly identical in the 2D and 3D case. The peak of the entropy distribution is found near zero (dashed grey line), which is the adiabatic entropy. The distribution is skewed to the left, which indicates that the entropy fluctuation of the cold downflows are larger in magnitude on average than the hot upflows. In contrast the PDFs of vertical velocity differ substantially between 2D and 3D. The asymmetry between upflows and downflows is visible in the vertical velocity distribution, where the upflows on the right side of the dashed grey line (zero velocity) are closer on average to zero than the downflows on the left hand side. The 2D and 3D cases have distinct velocity distributions. The 3D case is a narrower distribution overall with a kurtosis of 7.52 as compared to 4.54 for the 2D case, indicating that the velocities are smaller overall, and the skew towards negative velocity is more pronounced, with skewness of -1.74 as compared to -0.60 for the 2D case. This likely is related to the unique dynamics created with flows restricted to a plane such as the formation of long-lived flywheel structures in 2D, but not in 3D.

\subsection{Structure}
\begin{figure}
    \centering
    \includegraphics[width=\textwidth]{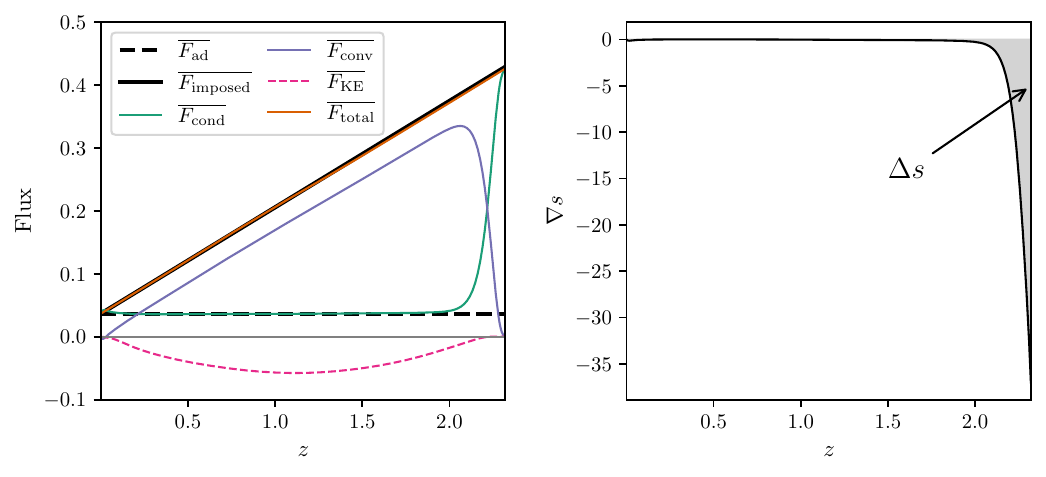}
    \caption{(Left panel) Time- and horizontally-averaged vertical energy flux profiles for the 3D simulation with $\mathrm{Ra}/\mathrm{Ra}_\mathrm{crit}=2\times10^4 $ and $\Heat=10.667$ plotted against height. The imposed flux from the source term is shown with the solid black line and the adiabatic flux (defined in equation \ref{eqn:f_ad}) with the dashed black line. The time-averaged total flux $\overline{F_\mathrm{total}}$ is shown with the orange line which converges to imposed heat flux $\overline{F_\mathrm{imposed}}$. The green line shows the conductive flux $\overline{F_\mathrm{cond}}$ which is adiabatic at the lower boundary and interior. The conductive flux becomes large in the superadiabatic upper boundary layer. We show the convective flux $\overline{F_\mathrm{conv}}$ with the purple line. We also display the kinetic energy flux, a component of the convective flux, with a dashed pink line. Due to the asymmetry introduced by stratification the KE flux is negative. In the right panel we show the entropy gradient. The entropy gradient is zero everywhere except the superadiabatic upper boundary. The integral of the entropy gradient, denoted by the shaded grey region, is the change in entropy, $\Delta s$, which is important in predicting the Mach number of the flows as we discuss in section \ref{subsec:delta_s}.}
    \label{fig:profiles}
\end{figure}

    The downflow-dominated dynamics can be explained by horizontally averaged profiles of the energy fluxes and the entropy gradient, which are shown in Fig. \ref{fig:profiles}. In the left panel, we show the vertical component of the horizontally and time averaged fluxes. We denote a horizontally averaged quantity with a bar, ie $\overline{a} = \frac{1}{L_x L_y} \int a\,dx\,dy $ where $L_y$ and $\partial y$ are dropped for 2D simulations. To derive the fluxes for this system, we first define a total energy equation by summing  $\rho \vec{u}$ dotted into Eqn. \ref{eqn:momentum} and $\rho c_V$ multiplied into \ref{eqn:temperature}. After invoking Eqn. \ref{eqn:continuity} and some manipulation we find
    \begin{equation}
        \frac{\partial}{\partial t}\left(\frac{1}{2}\rho|\vec{u}|^2 +\rho( -g z) + c_V\rho T \right) 
        + \nabla \cdot \left [ 
        \rho \vec{u} \left(\frac{1}{2} |\vec{u}|^2\right)
        + \rho\vec{u} (-g z)
        - \mu(\vec{u}\cdot \sigma)
        + \rho \vec{u} (c_P T) - 
        \kappa \nabla T \right]
        =Q.
    \end{equation}
This equation is in conservation form with the sum of the kinetic energy, potential energy, and internal energies in the time derivative, the sum of the fluxes in the divergence term, and our source term Q on the right hand side.
We define the conductive, convective, and total fluxes as

    \begin{gather}
        \vec{F}_\text{cond} = -\kappa \grad T \qquad\mathrm{(conductive\,flux)} \label{eqn:f_cond}
 \\
        \vec{F}_\text{conv} = \vec{F}_\text{enth} + \vec{F}_\text{KE} +  \vec{F}_\text{PE} +  \vec{F}_\text{visc} \qquad\mathrm{(convective\,flux)}  \label{eqn:f_conv}
 \\   
        \vec{F}_\text{total} = \vec{F}_\text{cond} + \vec{F}_\text{conv} \qquad\mathrm{(total\,flux)}.  \label{eqn:f_totla}
    \end{gather}
   The components of the convective flux are the enthalpy flux, kinetic energy flux, and potential energy flux:
    \begin{gather}
       \vec{F}_\text{enth} = \rho \vec{u} (c_P T)\qquad\mathrm{(enthalpy\,flux)}  \label{eqn:f_enth}\\
       \vec{F}_\text{KE}=\rho \vec{u} (|\vec{u}|^2/2) 
       \qquad\mathrm{(kinetic\, energy\,flux)} \label{eqn:f_ke}\\
       \vec{F}_\text{PE} = \rho \vec{u} (-g z)  \qquad\mathrm{(potential\, energy\,flux)} \label{eqn:f_pe}\\
       \vec{F}_\text{visc} =- \mu (\mathbf{u}\cdot\sigma) \qquad\mathrm{(viscous\,flux)}.  \label{eqn:visc}
    \end{gather}
    We further define the adiabatic flux, which is the flux conducted along the adiabatic temperature gradient, and the flux imposed by our heat source which is defined such that our fixed flux lower boundary is satisfied as 
    \begin{gather}
        \vec{F}_{\rm ad} = -\kappa \grad T_{\rm ad} = -\kappa \vec{g}/c_P \qquad\mathrm{(adiabatic\,flux)} \label{eqn:f_ad}\\
        \vec{F}_\mathrm{imposed} = Q z +\vec{F}_\mathrm{ad}\qquad\mathrm{(imposed\,flux)}.  \label{eqn:f_q}
    \end{gather}
    Our systems are in thermal equilibrium, denoted by the fact that on average the total flux through the system matches $\vec{F}_\mathrm{imposed}$. In the interior, conduction transports a flux equivalent to the adiabatic flux, and convection efficiently transports the remainder of the flux. The kinetic energy flux (dashed pink line) is small and directed downward, thus the enthalpy flux is large and positive. In the upper boundary layer convective motions come to a stop to satisfy the impenetrable boundary conditions and the dominant form of heat transport transitions to conduction giving rise to a superadiabatic upper boundary layer (blue layer visible in Fig. \ref{fig:dynamics} at top boundary). 
    
    These same features are seen in the right panel of Fig. \ref{fig:profiles} where we plot the specific entropy gradient $\grad s$, where
    \begin{equation}
            s=(\gamma-1)\ln T - \ln \rho.
    \end{equation}
    The entropy gradient follows the adiabatic value in the bulk but becomes strongly superadiabatic near the upper boundary. Strong downflow plumes launch from this superadiabatic boundary layer. Note the lack of a lower boundary layer, which is consistent with the absence of strong upflow plumes in Fig. \ref{fig:dynamics}. The size of the upper boundary layer can be measured by the integral of the entropy gradient, $\Delta s = \int \partial s \partial z = s(z=L_z)-s(z=0)$, which scales with both $\Heat$ (which varies the magnitude of the boundary layer), and Ra (which varies the width of the boundary layer). We will discuss the scaling of $\Delta s$ with our control parameters as well as its impact on the Mach number in section \ref{subsec:delta_s}.

\subsection{Scalings}
\label{subsec:scalings}
\begin{figure}
    \centering
    \includegraphics[width=\textwidth]{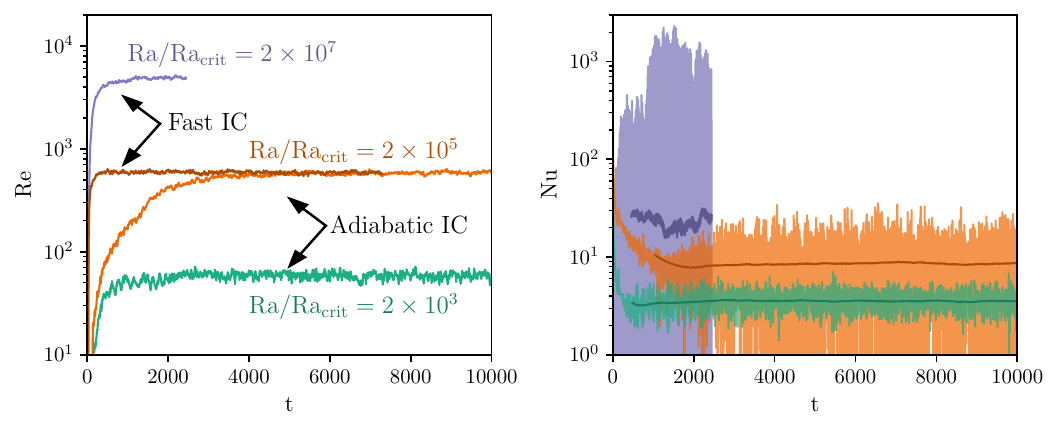}
    \caption{Time evolution of Reynolds number (left panel) and Nusselt number (right panel). 
    We display time traces for three 2D simulations at $\Heat=0.75$ at $\mathrm{Ra} /\mathrm{Ra}_\mathrm{crit} = 1.9\times 10^3,  1.9\times 10^5 \,\mathrm{and}\,  1.9\times 10^7$. The $\mathrm{Ra} /\mathrm{Ra}_\mathrm{crit} = 1.9\times 10^7$ simulation uses fast-IC, and the $\mathrm{Ra} /\mathrm{Ra}_\mathrm{crit} = 1.9\times 10^3$ simulation uses an adiabatic polytrope for its initial condition. We show both adiabatic (orange line) and fast initial conditions (brown line) for the simulation at $\mathrm{Ra} /\mathrm{Ra}_\mathrm{crit} = 1.9\times 10^5$ in the Reynolds number plot.  The simulations using fast-IC reach a thermally converged state quickly. At high Ra, the instantaneous value of Nu becomes chaotic, so we also plot the rolling time average over 40 heating timescales which is denoted by the darker colored trace.}
    \label{fig:traces}
\end{figure}

We calculate time and volume averaged Mach number, Reynolds number, and Nusselt numbers, which we respectively define as
\begin{equation}
    \text{Ma} = \angles{\frac{\sqrt{|\vec{u}|^2}}{c_s}},\qquad
    \text{Re} = \angles{\frac{\sqrt{|\vec{u}|^2}L_c}{\nu}},\qquad
    \text{Nu} = 1+\angles{\frac{F_\text{conv}}{\angles{F_\text{cond} - F_\text{ad}}}} = \angles{\frac{F_\text{Q}}{\angles{F_\text{cond} - F_\text{ad}}}} .
    \label{eqn:nondim_outputs}
\end{equation}
Here, $\angles{a} = \frac{1}{L_x L_y L_z }\int a \, dx \, dy \, dz$ represents a volume-average (where we drop $L_y$ and $\partial y$ for 2D simulations), and we only use the vertical components of the fluxes (defined in equations \ref{eqn:f_cond} - \ref{eqn:f_q}) to calculate Nu. Our Nusselt number takes the same form as the compressible Nusselt number used in \citet{Graham1975JFM, Hurlburt1984ApJ...282..557H} and \citet{AndersBrown2017}. This Nusselt number is the ratio of the total flux in excess of the adiabatic flux divided by the conductive component of the flux in excess of the adiabat. As we approach the Boussinesq limit, $F_\mathrm{ad}$ approaches zero, and we recover the standard Boussinesq Nusselt number.

The time evolution of Re and Nu are displayed in Fig. \ref{fig:traces}. We show time traces from three 2D simulations at $\Heat=0.75$ with different values of Rayleigh number. The highest Rayleigh number run (purple line) uses Fast-IC, and it reaches a thermally equilibriated solution at approximately $t=1000$ whereas the lower Rayleigh number run at $\mathrm{Ra}/\mathrm{Ra}_\mathrm{crit} = 2\times 10^5$ (orange line) takes until approximately $t=2500$ despite having a shorter diffusive timescale. These simulations converge to a statistically similar state. We find that the simulation shown in orange in Fig. \ref{fig:traces} reaches $\mathrm{Re}=582$ with the standard initial conditions and $\mathrm{Re}=589$ with ``Fast-IC''. The standard deviation of the fluctuations in Re are $\sigma=14$, so these two simulations reach values of Re that are well within one standard deviation. While fluctuations of Re are relatively small compared to the mean, the fluctuations of Nu are large and grow significantly with Rayleigh number, indicating that nonlinearity becomes more important. In this highly nonlinear state, such as the orange and purple trace in Fig. \ref{fig:traces}, minima in the Nusselt number are caused when cold fluid is swept upwards and hot fluid is swept downwards. Because of the high Peclet number ($\mathrm{Pe} \approx 5 \times 10^3$), these hot and cold fluid parcels retain their thermal signature for several domain crossings, as is seen in Refs. \citep{Johnston2009_temp_vs_flux_bc, Zhu2018_ultimate_regime, Anders2018}.
\begin{figure}
    \centering
    \includegraphics{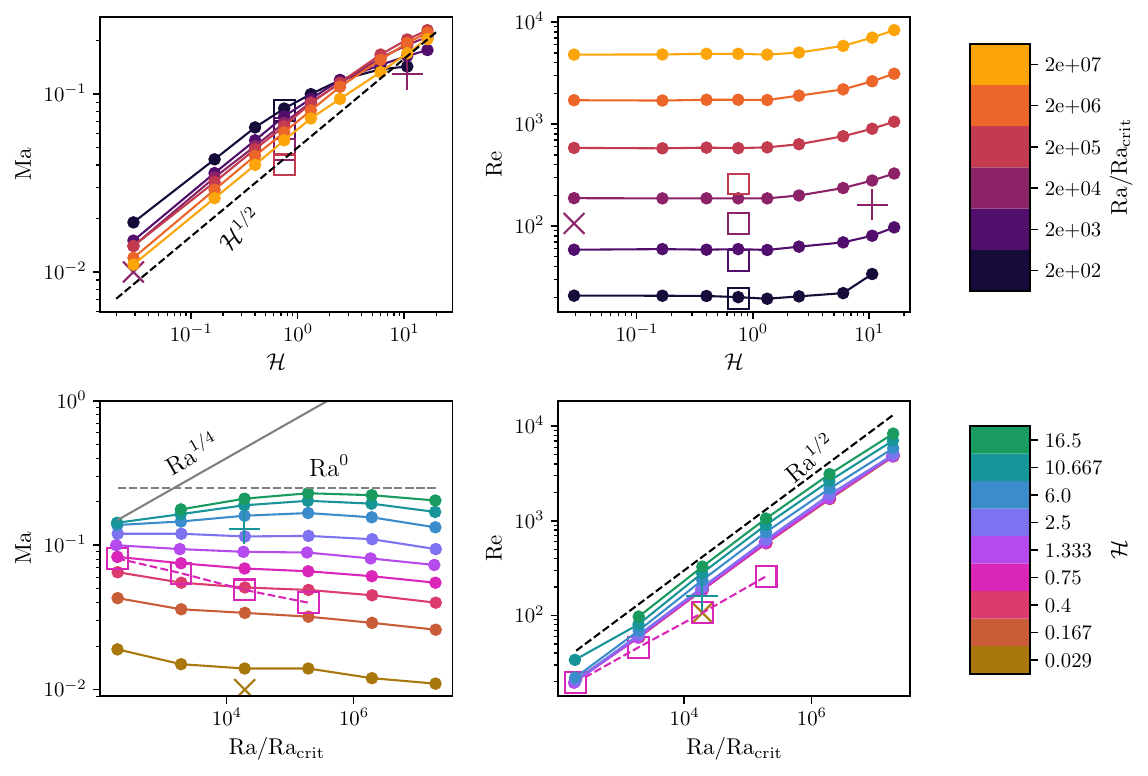}
    \caption{ (Upper left panel) The scaling of average Mach number against $\Heat$, where color represents $\mathrm{Ra}/\mathrm{Ra}_\mathrm{crit}$. Dots represent 2D simulations and the $\times$, box and $+$ symbols represent 3D simulations. We display a $\Heat^{1/2}$ power law with a dashed black line which approximates the scaling observed in the data. (Upper right panel) The scaling of average Reynolds number against $\Heat$, where color represents $\mathrm{Ra}/\mathrm{Ra}_\mathrm{crit}$. (Bottom left panel) The scaling of Mach number with Rayleigh number where color denotes $\Heat$. Note that in 3D there is a downward trend in Mach number with increasing Rayleigh number. The power laws found by Ref. \citep{AndersBrown2017} (AB17) are shown by grey lines, with the solid line showing the 2D power law from AB17 and the dashed line showing the 3D power law. (Lower right panel) the scaling of average Reynolds number against supercriticality where color denotes $\Heat$. We show a $\mathrm{Ra}^{1/2}$  power law with a dashed black line.}
    \label{fig:scaling}
\end{figure}

 In the upper left panel of Fig. \ref{fig:scaling}, we show the scaling of average Ma against $\Heat$. We find a scaling that is roughly consistent with a $\Heat^{1/2}$ power law at $\Heat \lesssim 1$ and a weaker scaling at large $\Heat$. This transition at $\Heat \sim 1$ is where $\nabla T_1(z=L_z)\sim\nabla T_{ad}$  The scaling deviates from the $\Heat^{1/2}$ power law more rapidly at lower Ra. While we do not have results from higher values of $\Heat$, and so we cannot be conclusive, it appears that in the high $\Heat$, higher Mach number limit, Mach number may become constant as $\Heat$ increases. In the upper right panel, we plot Re vs.~ $\Heat$. Re is independent of $\Heat$ when $\Heat \lesssim 6$. Re increases with $\Heat$ as we transition to higher Mach number flows at $\Heat \gtrsim 6$. While we start to see some scaling of Re with $\Heat$, the Reynolds number still scales like $\mathrm{Ra}^{1/2}$ in this regime. In the lower left panel we show the scaling of average Ma against supercriticality. We see that our 2D simulations have a roughly constant Mach number for a fixed value of $\Heat$, however Mach number decreases with Ra for our 3D simulations. In the lower right panel we plot average Reynolds number against supercriticality. Our simulations roughly match a $\mathrm{Re} \propto \mathrm{Ra}^{1/2}$ power law which is the expected behavior.  Re scales weakly with $\Heat$, so we achieve more turbulent flows at fixed supercriticality. 

 We perform a least-squares fit for Re and Ma for 2D simulations with $\Heat < 6$. We exclude values at higher $\Heat$ due to the lack of clean power law behavior in Ma as can be seen in Fig. \ref{fig:scaling} (e.g., top left panel).  We find 
\begin{equation}
    \text{Ma}_{2\mathrm{D}} \propto \left(\frac{\text{Ra}}{\text{Ra}_\text{crit}}\right)^{-0.02} \Heat^{0.44}.
\end{equation}
The Reynolds number scales as 
\begin{equation}
\text{Re}_{2\mathrm{D}} \propto \left(\frac{\text{Ra}}{\text{Ra}_\text{crit}}\right)^{0.45} \Heat^{0.01},
\end{equation}
 Importantly for 2D simulations, we find little Re dependence on $\Heat$, and little Ma dependance on Ra. This validates our experimental design. We performed a least squares fit for the scaling or Re and Ma against Ra for the suite of 3D simulation with constant $\Heat =0.75$. We find that while the 2D simulations produce roughly constant Ma for a given value of $\Heat$,  Ma decreases with Ra in 3D as $\mathrm{Ma}\propto\mathrm{Ra}^{-0.10}$.
Additionally we find that the 3D simulations achieve a weaker Re scaling of $\mathrm{Re} \propto \mathrm{Ra}^{0.38}$ than our 2D simulations.

\begin{figure}
    \centering
    \includegraphics[width=0.625\textwidth]{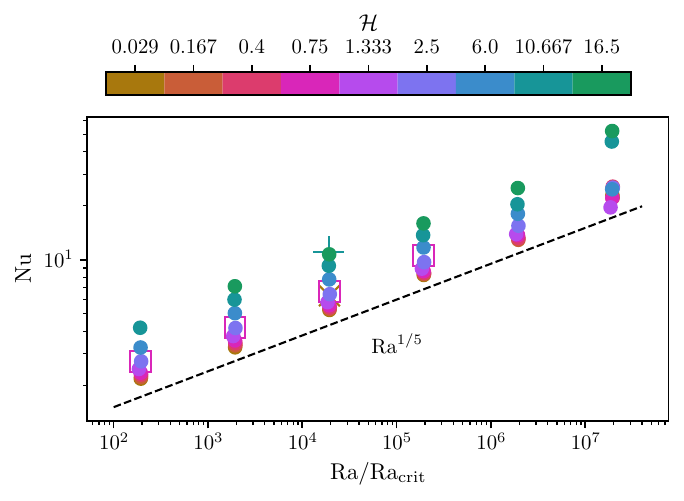}
    \caption{ We show Nusselt number against Rayleigh number for our suite of simulations. We find a $\text{Nu} \propto (\text{Ra}/\text{Ra}_\text{crit})^{1/5}$ scaling, which is consistent with the incompressible internally heated simulations from \citet{goluskin_spiegel_2012}.}
    \label{fig:Nu_scaling}
\end{figure}

The Nusselt number is a measure of how efficiently convection transports heat. It is expected that the Nusselt number in astrophysical convection follows a diffusion-free power-law scaling which is referred to as the ``ultimate regime'' \citep{Spiegel1963}. We measure a Nusselt number scaling law using our input Rayleigh number, and given this choice we expect the ``ultimate regime'' to be a $\mathrm{Nu}\propto\mathrm{Ra}^{1/3}$ power law. We measure the scaling of the Nusselt number with $\text{Ra}/\text{Ra}_\text{crit}$ and show these results in Fig. \ref{fig:Nu_scaling}. We find  $\mathrm{Nu} \propto (\text{Ra}/\text{Ra}_\text{crit})^{0.21}$ for 2D and $\mathrm{Nu} \propto (\text{Ra}/\text{Ra}_\text{crit})^{0.20}$ for 3D, both are very close to a $1/5$ power law. As we transition to a higher Mach number regime at the upper range of $\Heat$ we find that the Nusselt number increases with $\Heat$ and that a $1/5$ power law does not fit our data as effectively at high $\Heat$ where we observe a steeper trend at high Ra. This suggests that compressible effects modify the boundary layer formation and scaling for high Mach number convection. 

\subsection{Effect of Upper Boundary Layer}
\label{subsec:delta_s}
To study the origin of our observed scaling of Mach number with $\Heat$ we measure the change in specific entropy, $s$, across the domain.  Since the entropy profile is adiabatic everywhere except the upper boundary, this domain-average measures the properties of the upper boundary layer. $\Delta s$ is a function of both $\Heat$ and Ra, where increasing $\Heat$ increases the magnitude the boundary layer, and increasing Ra decreases its width. In the low-Mach number regime where the anelastic approximation is valid, a balance between buoyancy and nonlinear inertia takes the form \citep{brown_etal_2012}
\begin{equation}
    \vec{u}\cdot\grad\vec{u} \sim -\frac{\vec{g}}{c_p} s
    \Rightarrow
    u \sim \sqrt{\ell \frac{g}{c_p} s},
    \label{eqn:balance}
\end{equation}
where $\ell$ is a dominant flow length scale. This suggests that we should find $\mathrm{Ma}\sim (\Delta s)^{1/2}$.
    
\begin{figure}
    \centering
    \includegraphics[width=\textwidth]{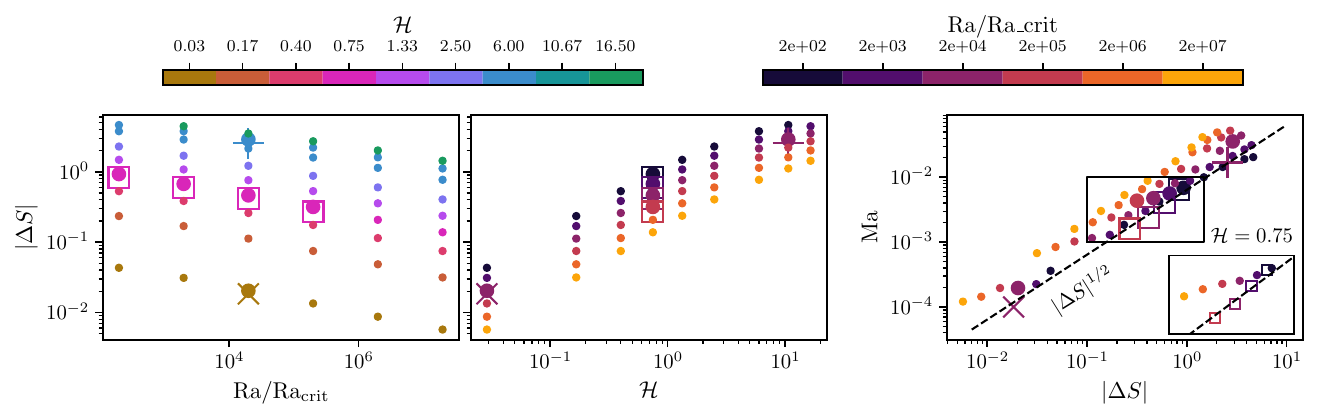}
    \caption{In the left panel we display a scatter plot of $\Delta s$ against $\text{Ra}/\text{Ra}_\text{crit}$. Colors correspond to values of $\Heat$, and squares, x, and + symbols correspond to 3D runs. Circles denote 2D runs, with large circles indicating a 2D run with a corresponding 3D run at the same control parameters.  In the middle panel we show a scatter plot of $\Delta s$ against $\Heat$, with the same marker styling as in the left panel, but with colors now corresponding to Rayleigh number. In these two panels we show that the scaling of $\Delta s$ with our control parameters is consistent between 2D and 3D simulations and the small differences that are seen are not from a change in the power law exponent. In the right panel we show how Mach number scales with $\Delta s$. We find a scaling consistent with $\text{Ma} \propto (\Delta s)^{1/2}$ for 3D simulations and for 2D simulations at a fixed value of Ra. However, when we increase Ra in 2D simulations $\Delta s$ decreases. We show an inlay plot with just the simulations with $\Heat=0.75$. We see that for 3D, all simulations fall on the $\text{Ma} \propto (\Delta s)^{1/2}$ trend line, but the 2D simulations do not, with increased Ra leading to smaller $\Delta S$ and slightly decreased Ma.
} 
    \label{fig:ma_delta_s}
\end{figure}

In Fig. \ref{fig:ma_delta_s} we plot the behavior of $\Delta s$ against our control parameters Ra (left panel) and $\Heat$ (middle panel). Increasing Rayleigh number causes the superadiabatic upper boundary to be thinner, causing a negative scaling, while increasing $\Heat$ causes the amplitude of the superadiabatic boundary to increase, casuing a positive scaling. The scaling of $\Delta s$ is similar for both 2D simulations (circles) and 3D simulations (squares, $\times$, and $+$ symbols). 3D simulations have slightly smaller boundary layers and thus smaller values $\Delta s$, but the scaling exponent remains the same. This indicates that the size of the boundary layer scales in a consistent manner for both 2D and 3D simulations. 

In the right panel of Fig. \ref{fig:ma_delta_s} we plot Ma against $\Delta s$. We find that our simulations are broadly consistent with a $\text{Ma} \propto (\Delta s)^{1/2}$ power law (dashed black line) suggesting that this is the dominant nonlinear trend driving our dynamics. Our 3D simulations are well described by a $(\Delta s)^{1/2}$ power law, but the 2D simulations have more deviation from the trend line indicating the presence of a second order effect from our control parameters on the Mach number of the flows. To better show this effect we plot only the simulations with $\Heat=0.75$ in the inlay. We see that the 3D simulations all fall on the $(\Delta s)^{1/2}$ trend line, however the 2D simulations depart from this trend line where we find that higher Ra simulations produce higher Ma flows than predicted from the anelastic theory. This discrepancy between 2D and 3D simulations cannot be explained by the size of the boundary layer. Instead the way in which the dynamics respond to the boundary layer changes between the 2D and 3D case. Since all the 3D simulations fall on a single trend line, it should be possible to design 3D simulations that reach constant Ma against variations in supercriticality as long as the change in entropy is held constant.

\subsection{Power Spectra}
We calculate the convective power spectra at the mid-z plane of our 2D simulations at $\Heat=0.75$. To do so, We output the interpolated velocity at $z=L_z/2$, and compute the discrete Fourier transform. We define the Fourier transform of the velocity as 
\begin{equation}
    \mathbf{U}(k_x) = \sum_{n=0}^{N_x-1}\vec{u}(x_n)e^{-\frac{i2\pi}{N_x}k_x n}
\end{equation}
We define the power spectral density as the product of the Fourier transform of velocity and its complex conjugate normalized by the horizontal resolution squared, $\mathcal{P}(k_x) = (1/N_x^2) \mathbf{U}(k_x)^*\cdot\mathbf{U}(k_x)$, where $*$ denotes the complex conjugate and $N_x$ is the horizontal resolution.

For 3D simulations we perform a 2D Fourier transform of the velocity at $z=L_z/2$ for simulations at $\Heat=0.75$. We define the 2D discrete Fourier transform as 
\begin{equation}
     \mathbf{U}(k_x, k_y) = \sum_{n=0}^{N_x-1} \sum_{m=0}^{N_y-1} \vec{u}(x_n, y_m)e^{-i2\pi \left(\frac{k_x n}{N_x} +\frac{k_y m}{N_y}\right) }
\end{equation}
We define the power spectral density as $\mathcal{P}(k_x, k_y) = (1/(N_x^2 N_y^2)) \mathbf{U}(k_x, k_y)^*\cdot\mathbf{U}(k_x, k_y)$ and then transform the power spectra density from $k_x$ and $k_y$ space to $k_\theta$ and $k_h$ space where $k_h = \sqrt{k_x^2+k_y^2}$ is the horizontal wavenumber and $k_\theta = \mathrm{arctan}(k_y/k_x)$ we then interpolate and average over $k_\theta$ to produce a horizontal wavenumber power spectral density $\mathcal{P}(k_h)$.

For 2D simulations at low wavenumber we find that the power spectrum magnitude changes little as Ra is increased. We expect this behavior in 2D since Mach number and Reynolds number are independent of each other under our choice of parameterization. The spectra decay with  roughly a $k^{-3}$ power law, which is associated with a two dimensional turbulent cascade of vorticity from low to high wavenumber \citep{Kraichnan1967TurbulentCascades, ChenGlatzmaier2005}. The characteristic scale of our internal heat source is the size of the domain, so it makes sense that we find the $k^{-3}$ forward vorticity cascade as we are driving convection over a large length scale. At high $k_x$ we see the viscous cutoff wavenumber increase with Ra, but the nature of the spectra at lower $k_x$ remains unchanged. This suggests that it may be possible to extrapolate the large-scale dynamics present in our simulations to higher Rayleigh number cases. However, the large scales of our 2D simulations are dominated by flywheel structures, which are not found in astrophysical convection, limiting the utility of this behavior.

We find that the spectra of our 3D simulations does not have the same consistent behavior as seen in 2D. As Rayleigh number increases, the magnitude of the low wavenumber modes decreases.  We expect this behavior from the 3D Mach number scaling since the magnitude of the power spectrum depends on the Mach number. This unfortunately means that less turbulent simulations cannot be easily used to predict low wavenumber flows in a more turbulent system.

\begin{figure}
    \centering
    \includegraphics[width=\textwidth]{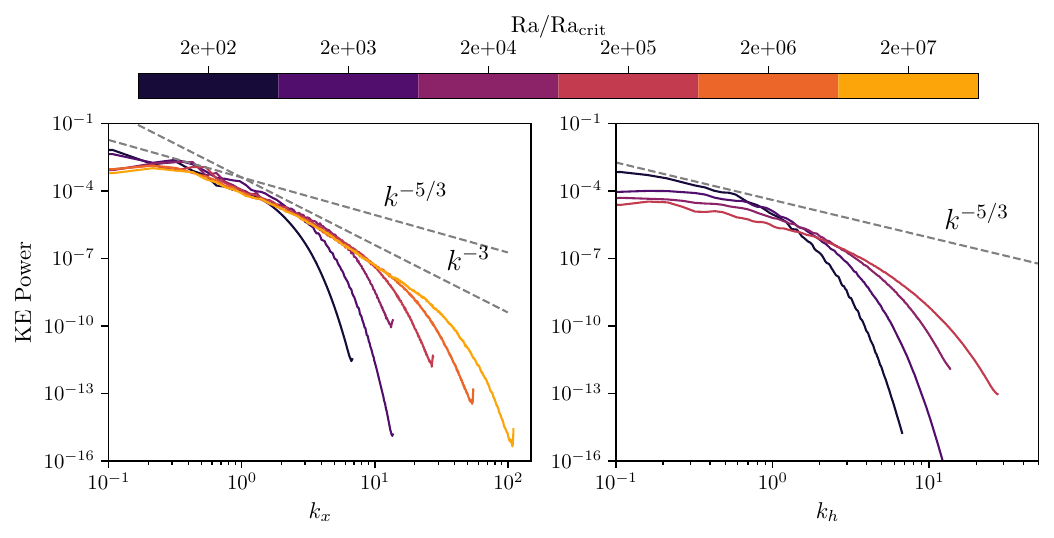}
    \caption{(left panel) Convective power spectra of 2D simulations accumulated over 40 heating timescales. We plot the power spectra of the mid-z plane velocity vector for a set of simulations with constant \Heat. We find consistent power at low wavenumber for these simulations. The less turbulent simulations enter the viscous regime at lower $k_x$ than the more turbulent simulations. We plot canonical $k^{-5/3}$ and $k^{-3}$ power laws, and note that the $k^{-3}$ power law better matches the energy cascade in our simulations. The sharp vertical cutoff of the spectra is found at the highest possible wavenumber given the simulation's resolution. (right panel) Convective power spectra of 3D simulations accumulated over 2 heating timescales. We plot power spectra of the mid-z plane velocity vector for a suite of simulations at constant $\Heat$. The low wavenumber power decreases with increasing Ra.} 
    \label{fig:spectra}
\end{figure}
\subsection{Comparison to Other Work}
It is useful to compare our results to simulations of fully compressible boundary driven convection, as well as incompressible internally heated convection. The former allows us to interpret the effect of different thermal driving models, while the latter allows us to study how compressibility affects simulations with otherwise similar thermal driving. Our dynamics are distinct from the characteristic behaviour of boundary driven convection. We observe dynamics with strong downflows and weak, diffuse upflows. In our 3D simulations downflows form individual plumes and are not pushed into downflow lane structures. This is counter to what is seen in a boundary driven system \citep[e.g.,][]{AndersBrown2017} The downflow dominated dynamics we observe are reminiscent of the convective dynamics observed by \citet{kapyla_etal_2017}, who studied a depth-dependent internal heating, despite the fact that we study a constant heat source.

We measure scalings of Reynolds number, Mach number, and Nusselt number. Our Reynolds number scalings of $\mathrm{Re} \propto \mathrm{Ra}^{0.45}$ for 2D and $\mathrm{Re}\propto\mathrm{Ra}^{0.38}$ for 3D are similar to those in \citet{AndersBrown2017} (AB17) who found a $\mathrm{Re}\propto\mathrm{Ra}^{3/4}$ power law for 2D low Mach simulations and a $\mathrm{Re}\propto\mathrm{Ra}^{1/2}$ power law for 3D simulations and $\mathrm{Ma} \sim 1$ 2D simulations. However our scaling exponent in 3D is smaller. This similarity is unsurprising since our Rayleigh number scales viscosity and diffusivity in the same basic manner. Our Mach number scaling of $\mathrm{Ma}\propto \mathrm{Ra}^{-0.02}$ for 2D and $\mathrm{Ma}\propto \mathrm{Ra}^{-0.10}$ for 3D however is in stark contrast with the scalings found by AB17, where 2D simulations followed a $\mathrm{Ma}\propto\mathrm{Ra}^{1/4}$ power law and 3D simulations had a constant Mach number with a fixed superadiabatic excess ($\epsilon$) which is comparable to $\Heat$. Like AB17 we find stronger positive scalings in 2D than in 3D. It should also be noted that AB17 used fixed temperature boundary conditions at both boundaries whereas we use mixed flux-temperature boundary conditions.

Our Nusselt number scaling law of $\mathrm{Nu}\propto\mathrm{Ra}^{1/5}$ is comparable to the Nusselt number scalings of the boundary driven fully compressble convection simulations of AB17 in their high-Ra 2D simulations, however their low-Ra 2D simulations follow a 1/3 power law, and their 3D simulations follow a 2/7 power law. The 2/7 power law was also observed with fully compressible boundary driven convection by \citep{Johnston2009_temp_vs_flux_bc} and a steeper scaling exponent of 0.38 was observed by \citep{Zhu2018_ultimate_regime}. We can also compare our Nusselt number scaling to incompressible internally heated convection studies. We find the same Nusselt number scaling as \citep{goluskin_spiegel_2012,Goluskin_vanderPoel_2016JFM}, however their simulations used fixed temperature at the upper and lower boundaries. Therefore, the uniform heating case of \citep{kazemi_etal_2021} is the more analogous incompressible study for our setup. Note that \citep{Goluskin_vanderPoel_2016JFM, kazemi_etal_2021} report a scaling law based on a ``diagnostic" Rayleigh number which is equivalent to $\mathrm{Ra}/\mathrm{Nu}$. We have converted their results to an input Rayleigh number scaling to be consistent with our choice of Ra and Nu.  \citet{kazemi_etal_2021} found a $\mathrm{Nu}\propto\mathrm{Ra}^{1/4}$ scaling, which is close to our scaling law, but not identical. It is striking that we find a similar Nusselt number scaling to \citep{kazemi_etal_2021}, and a near identical scaling to \citep{goluskin_spiegel_2012, Goluskin_vanderPoel_2016JFM} despite the different boundary conditions.
\section{Conclusions}
\label{sec:conclusions}
In this work we simulated fully compressible, internally heated convection. We used 2D and 3D simulations and we varied the heating, $\mathcal{H}$, and Rayleigh number, Ra. We observed convection dominated by cold downflows. We did not observe the formation of hot upflow plumes at the lower boundary. This agrees with our expectations as the lower boundary is thermodynamically stable. The dynamics we observe appear similar to those produced by \citet{cossette_rast_2016, kapyla_etal_2017}, where the downflows are buoyantly driven and the upflows are pressure driven and heated as they rise. 

We found that the Mach and Reynolds numbers could be independently controlled in 2D by using $\Heat$ and Rayleigh number as input parameters. Varying Ra produces a $\mathrm{Re} \propto \mathrm{Ra}^{1/2}$ and a weak Ma scaling. Varying $\Heat$ produces a $\mathrm{Ma} \propto \Heat^{1/2}$ scaling and a weak Re scaling. In 3D we find a weaker scaling for Reynolds number of $\mathrm{Re}\sim\mathrm{Ra}^{1/3}$, and a negative Ra scaling for Mach number of $\mathrm{Ma}\sim\mathrm{Ra}^{-1/10}$.

We found that the entropy jump in the upper boundary layer determines the average Mach number of the flow, and thus the magnitude of the kinetic energy power spectrum. While the change in entropy across the boundary layer scales similarly in 2D and 3D simulations, the impact of the boundary layer on the Mach number differs, with 2D simulations having an additional effect tied to the Rayleigh number. This likely reflects flywheel modes since these make up the large scale motions of our 2D simulations.

More robust mechanisms for controlling convective driving, such as an internal cooling layer which does not vary in size between experiments, could improve our ability to separate the resultant Re and Ma of the flows in 3D, and thus the behavior of power spectra. Internal heating and cooling layers have been used to reduce the effect of the boundaries on heat transport in simulations \citep{kazemi_etal_2021, barker_etal_2014} and experiments \citep{Lepot2018IntHeatExp, Bouillaut2019IHCExperiment} of Boussinesq convection. Since we find consistent heat transport scaling with the analogous internally heated Boussinesq setup of Goluskin \& Spiegel \citep{goluskin_spiegel_2012}, we expect that the use of a cooling layer may have similar effects in the fully compressible regime.  Critically, however, the relationship between a cooling layer width and heating magnitude may differ. Therefore further study on the shape of a non-uniform heating function is needed.

We additionally find heat transport scaling laws which are consistent with internally heated Boussinesq convection \citep{goluskin_spiegel_2012, Goluskin_vanderPoel_2016JFM, kazemi_etal_2021}. This suggests that certain properties of incompressible flows also hold for stratified flows with low Mach number and that studies of incompressible flows \citep[e.g.,][]{goluskin_spiegel_2012, Goluskin2015, kazemi_etal_2021, Bouillaut2019IHCExperiment} are useful for understanding astrophysical convection. However, this similarity may be limited to the low to intermediate Mach number regime which we studied in this work ($\mathrm{Ma} \leq 0.2$).

We also found that velocity power spectra of 2D simulations were invariant at low wavenumber while turbulence (Re) increased when the Mach number was held constant. This behavior may be useful for studying large scale flows since their properties do not change significantly as Ra increases, thus allowing us to study them with low resolution, low Ra simulations. However, with 3D simulations, the power at low wavenumber decreases as Ra increases. We may be able to achieve consistent low wavenumber behavior in 3D by using a non-uniform heat source with heating and cooling layers, as the properties of the flow would not be set by a thermal diffusion dominated upper boundary. 

Stars and gas giant planets do not have boundaries which act like hard walls, so the no-slip boundaries we used are likely a poor model for astrophysical flows and stress-free boundary conditions are likely more appropriate. We choose no-slip boundary conditions to prevent the onset of parasitic mean shear flows in 2D simulations. Mean shear flows reduce thermal transport by convection and suppress convective flows with predator-prey like dynamics \citep{Goluskin2014, Fuentes2021}. Wider aspect ratios have been shown to suppress shear flows in 2D simulations with stress-free boundaries \citet{wang_chong_stevens_verzicco_lohse_2020}, however this approach requires higher resolutions and longer convergence times. Shearing states typically do not emerge in 3D simulations, so future work relying exclusively on 3D simulations could utilize stress-free boundary conditions without this concern.

In this work we studied flows at Mach numbers between $10^{-2}$ and $0.2$. We find that the behavior of the Reynolds number and Nusselt number in our highest Mach number simulations deviated from the behavior at lower Mach number. Scaling laws for high Mach number flows could be inconsistent with the heat transport scaling from internally heated Boussinesq convection, and future studies should investigate this regime of flow. Future work should also study how internal heating interacts with other complications such as rotation \citep{Anders2019,Aurnou2020rotation} and chemical  mixing \citep{Bordwell}.

\acknowledgments
We thank Jeff Oishi, Daniel Lecoanet, and Bhishek Manek for insightful conversations about stratified convection. This work was supported by NASA HTMS grant 80NSSC20K1280 and NASA SSW grant 80NSSC19K0026. Computations were conducted with support from the NASA High End Computing (HEC) Program through the NASA Advanced Supercomputing (NAS) Division at Ames Research Center on Pleiades with allocations GID s2276 and s2114. W. Powers is supported through a Hale graduate research fellowship at the University of Colorado Boulder and a NASA FINESST Fellowship (grant number 80NSSC22K1850). E.~Anders  was supported by a CIERA Postdoctoral Fellowship.

\appendix

\section{Linear Stability Onset Analysis}
\label{app:ra_crit}
\begin{figure}
    \centering
    \includegraphics[width=\textwidth]{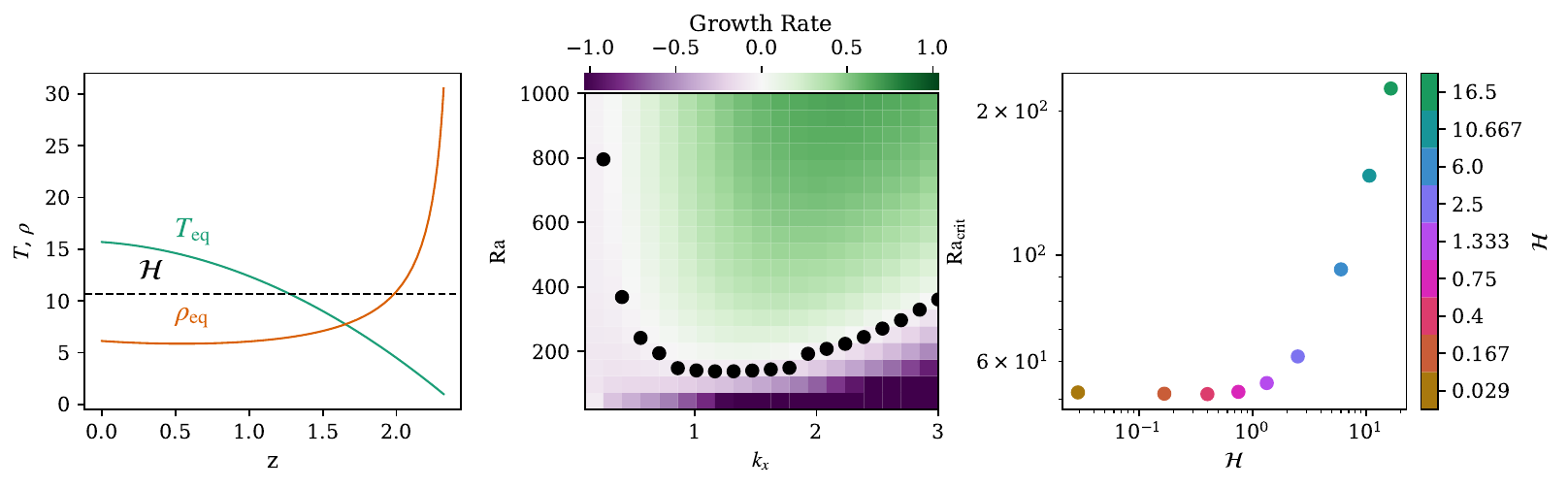}
    \caption{Left panel: Example of a thermal equilibrium solution with $\Heat = 10.667$ to be used as input to the linear stability analysis. We plot temperature, $T$ (green) and density, $\rho$ (orange), against height, $z$. Middle panel: Example of growth rates with $\Heat = 10.667$ plotted against horizontal wavenumber on the x-axis and Rayleigh number on the y-axis. Black dots indicate the interpolated value of Ra where the growth rate is zero for the given wavenumber. Right panel: Values of $\text{Ra}_\text{crit}$ plotted against $\Heat$. We find that $\text{Ra}_\text{crit}$ is roughly constant below $\Heat \sim 1$, and increases for $\Heat \gtrsim 1$.}
    \label{fig:apdxA}
\end{figure}

To determine the critical Rayleigh number $\text{Ra}_\text{crit}$ as a function of $\Heat$, we require an atmosphere in thermal and hydrostatic equilibrium. Our background reference state ($T_0$, $\rho_0$) is an adiabatic atmosphere, which satisfies hydrostatic equilibrium but is not in thermal equilibrium because of the internal heating. We solve for a stratification which satisfies thermal and hydrostatic equilibrium using a boundary value problem with an integral constraint to enforce mass conservation. We solve for
\begin{gather}
    -\frac{\partial^2 T_\text{eq}}{\partial z^2}=Q 
\\
    \frac{\partial \rho_\text{eq}}{\partial z} = -\frac{\rho_\text{eq}}{T_\text{eq}}\frac{\partial T_\text{eq}}{\partial z} - \frac{g \rho_\text{eq}}{R T_\text{eq}} \label{eqn:nlbpv_hse}
\end{gather}
where subscript eq denotes the equilibrium state. Our integral constraint is 
\begin{equation}
    \int_0^{L_z} \rho_\mathrm{eq} = \int_0^{L_z} \rho_0 \label{eqn:nlbpv_integral}
\end{equation}
where $\rho_0$ is our adiabatic reference state. The boundary conditions are
\begin{gather}
    \frac{\partial T_\text{eq}}{\partial z} (z=0) = \frac{\partial T_0}{\partial z}(z=0) = \grad T_\text{ad} \cdot \hat{z}
\\
    T_\text{eq}(z=L_z) = T_0(z=L_z) = (\grad T_\text{ad}\cdot \hat{z})L_z \label{eqn:nlbvp_bc_T_top}
\end{gather}
We show an example solution of this boundary value problem in the left panel of Fig \ref{fig:apdxA}.

We linearize Eqns.~\ref{eqn:continuity}-\ref{eqn:temperature} around this equilibrium state and perform a linear stability analysis using eigentools \citep{eigentools} to determine the onset of convection, taking solutions of form $e^{i \omega t}$ and searching for where the growth rate, $\omega$ reaches zero. We solve the boundary value problem and subsequent linear stability analysis for all values of $\Heat$. For each value of $\Heat$, eigentools calculates growth rates over a specified grid in Rayleigh number and horizontal wavenumber and then uses interpolation, root finding, and minimization algorithms to determine the minimum Rayleigh number where the growth rate is zero; see the middle panel of Fig \ref{fig:apdxA}. We show the resulting values of $\text{Ra}_\text{crit}$ in rhe right panel of Fig \ref{fig:apdxA}. We find that for $\Heat \lesssim 1$, Ra$_{\rm crit} \approx 50$, then grows as $\Heat$ increases.

\section{Fast Initial Conditions}
\label{sec:apdx_IC}
\label{app:initial_conditions}
To speed up the evolution to a thermally equilibrated, nonlinear convective state, we initialize our simulation with an approximation of the time averaged steady state (see Fig. \ref{fig:profiles}). We refer to this method as ``Fast-IC''. Convection efficiently mixes the temperature gradient to the adiabatic value in the interior. To satisfy thermal equilibrium and our boundary conditions, the time-averaged superadiabatic temperature gradient at the upper boundary must be $-\Heat$. We impose that the half-width of the thermal boundary layer is set by the Nusselt number as 
\begin{equation}
    \delta = \frac{L_z}{2 \text{Nu}}.
    \label{eqn:delta_nu}
\end{equation}
This scaling is chosen to approximate the width of a thermally equilibriated IVP solution. Our approximation of the steady state solution is 
\begin{equation}
    \frac{\partial T_{\rm eq}}{\partial z} = \grad T_\text{ad} \cdot \hat{z} + \frac{- \Heat}{2}\left[1+\text{erf}\left(\frac{\left(z-(L_z-\delta) \right)} { (\delta/2)}\right)\right]
    \label{eqn:T_smart_IC}
\end{equation}
where $\text{erf}$ is the error function. To ensure that the new initial conditions are in hydrostatic equilibrium and have the same mass as our original initial conditions, we solve a nonlinear boundary value problem to find $\ln\rho$ with equation \ref{eqn:nlbpv_hse} and integral constraint \ref{eqn:nlbpv_integral} while taking eqn. \ref{eqn:T_smart_IC} as $\partial T/\partial z$. $T_\text{eq}$ is obtained by integrating \ref{eqn:T_smart_IC} subject to boundary condition \ref{eqn:nlbvp_bc_T_top}.

After running our lower resolution 2D simulations with the adiabatic polytrope for initial conditions we fit a power law $\text{Nu} = A (\text{Ra}/\text{Ra}_\text{crit})^{\alpha}$ for each value of $\Heat$ with $A$ and $\alpha$ listed in Table \ref{tab:Nu_powerlaw}. We use this powerlaw to find the value of Nu in Eqn.~\ref{eqn:delta_nu} and then use the nonlinear boundary value problem to produce initial conditions for high Rayleigh number 2D runs and most 3D runs. We show an example of the specified shape for $T$ and the resulting solution for $\rho$ in Fig \ref{fig:roadmap_IC}.
\begin{table}
    \centering
    \begin{tabular}{c|c|c|c}
         $\Heat$ & $A$ & $\alpha$ & $\mathrm{Ra}_\mathrm{crit}$\\
         \hline
         0.029 & 0.7677 & 0.1943 & 51.58\\
         0.167 & 0.7920 & 0.1927 & 51.27\\
         0.400 & 0.8281 & 0.1902 & 51.16\\
         0.750 & 0.8571 & 0.1896 & 51.73\\
         1.333 & 0.9252 & 0.1868 & 53.97\\
         2.500 & 1.0270 & 0.1848 & 61.34 \\
         6.000 & 1.2251 & 0.1862 & 93.33 \\
         10.667 & 1.5539 & 0.1797 & 146.59 \\
         16.500 & 1.8926 & 0.1754 & 223.11
    \end{tabular}
    \caption{Parameters of the powerlaw fits for $\text{Nu} = A (\text{Ra}/\text{Ra}_\text{crit})^{\alpha}$}
    \label{tab:Nu_powerlaw}
\end{table}

\begin{figure}
    \centering
    \includegraphics{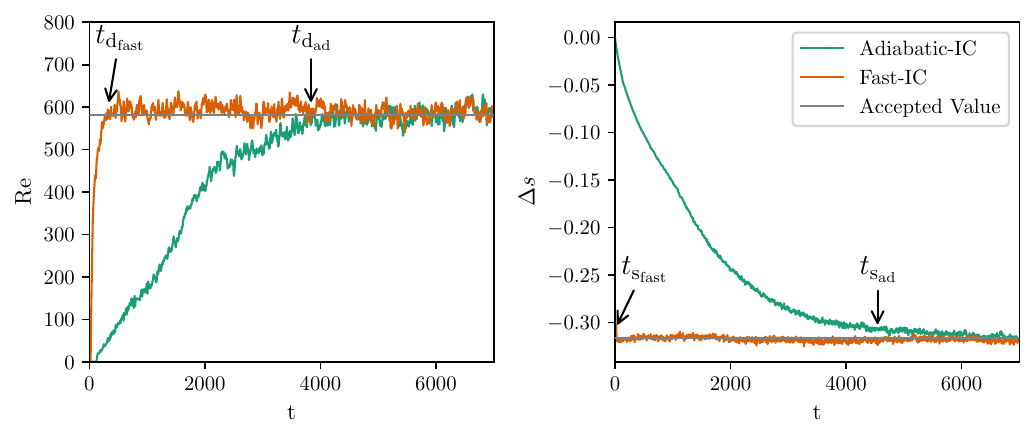}
    \caption{An example of time traces of Reynolds number (left panel) and $\Delta s$ (right panel) for simulations with $\mathrm{Ra} = 2\times 10^5 \mathrm{Ra}_\mathrm{crit}$ and $\Heat =0.75$ with ``Fast-IC" (orange) and adiabatic initial conditions(green). Grey horizontal lines show the time- and volume-averaged values reported in Appendix \ref{app:big_table}. Arrows show the time where a rolling average over 500 data points has converged to within $3\%$ of the the accepted value. }
    \label{fig:IC_traces}
\end{figure}

This method offers considerable reduction in CPU time required for running turbulent simulations in 2D. In Fig. \ref{fig:IC_traces} we show a comparison of the time-evolution of volume-averaged measures between two simulations with $\mathrm{Ra}/\mathrm{Ra}_\mathrm{crit}=2\times10^5$ and $\Heat=0.75$, one using ``Fast-IC", the other using adiabatic initial conditions. We plot (left) the Reynolds number, a measure of the dynamics, and (right) $\Delta s$, a measure of the structure. We calculate the time required for the Reynolds number to equilibrate to within $3\%$ of its final value. The dynamics convergence times are $t_{\mathrm{d}_\mathrm{fast}} = 320$ for ``Fast-IC" and  $t_{\mathrm{d}_\mathrm{ad}}=3800$ for adiabatic initial conditions. We calculate a $3\%$ convergence time for $\Delta s$ as well. We find that with ``Fast-IC" the $\Delta s$ begins within $3\%$ of the final value, and the structure convergence time for the adiabatic initial conditions is $t_{\mathrm{s}_\mathrm{ad}}=4500$. By predicting the final structure with ``Fast-IC", $\Delta s$ starts near its final value, whereas the adiabatic initial conditions slowly approach the converged state. ``Fast-IC" starts with $\vec{u}=0$, so the dynamics still take time to converge, however the time required is significantly shorter than with the adiabatic initial conditions.

\begin{figure*}[t!]
    \centering
    \includegraphics[width=\textwidth]{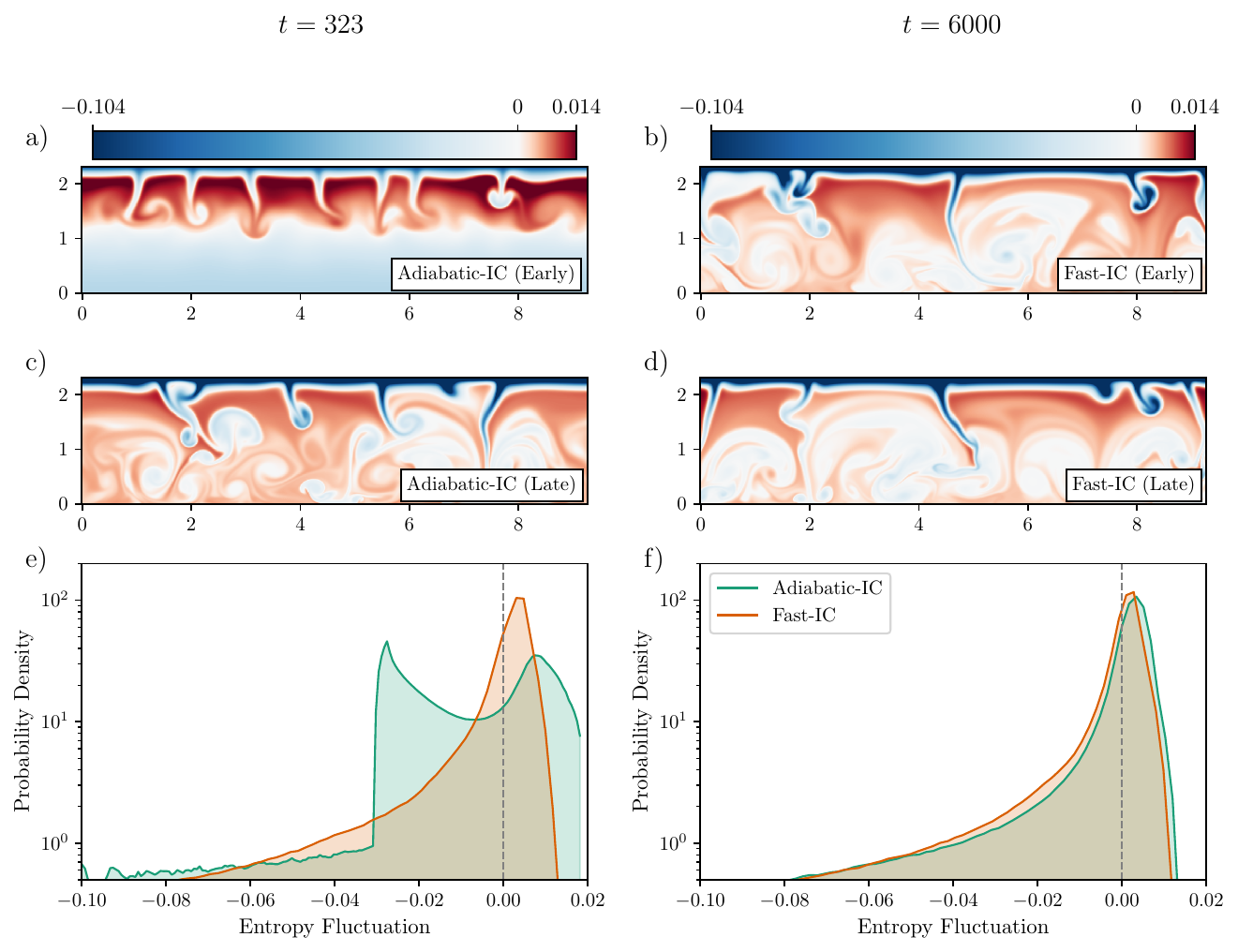}
    \caption{(Panels a and c) An example of entropy fluctuations at $ t=t_{\mathrm{d}_\mathrm{fast}}$ for simulations with $\mathrm{Ra} = 2\times 10^5 \mathrm{Ra}_\mathrm{crit}$ and $\Heat =0.75$ with adiabatic initial conditions (panel a) and ``Fast-IC" (panel c). The run with adiabatic initial conditions has reached convective onset, however cold downflow plumes stall out part way through the atmosphere, becoming more buoyant (redder) than the nearby fluid. As time progresses convective downflows reach further into the atmosphere. In the run with ``Fast-IC" the downflow plumes cross the domain immediatly after convective onset and the dynamics quickly reach a converged state. The colorbar is calculated from the adiabatic initial conditions case in the same manner as in Fig. \ref{fig:dynamics} and is used in all dynamics panels. In panel e we show probability density functions (PDFs) for entropy fluctuations averaged over 12 heating timescales for both adiabatic-IC and fast-IC simulations at $ t=t_{\mathrm{d}_\mathrm{fast}}$. The adiabatic-IC run has not converged at this time as evidenced by the discrepancy betweeen the PDFs. In the right column (panels b, d, and f) we show dynamics and PDFs at $t=6000$, which is after dynamics and structure have converged for both choices of initial conditions. The dynamics of the adiabatic-IC simulation (Panel b) are comparable to the dynamics of the Fast-IC run (Panel d). In panel f we show the PDFs for both runs where the distributions have converged, indicating that the dynamics have reached a statistically similar final state.}
    \label{fig:IC_dynamics}
\end{figure*}

In the left column of Fig. \ref{fig:IC_dynamics} we show the dynamics at $t_{\mathrm{d}_\mathrm{fast}}$ for both ``Fast-IC" and adiabatic initial conditions. With the adiabatic initial conditions in panel a the simulation has reached convective onset with cold plumes forming at the top of the domain.  The thermal boundary layer is shallow so the entropy perturbation of the plumes is small, therefore they loose their cool signature due to the heating before they transit the domain and halt partway down.  As the thermal structure evolves the plumes reach deeper into the domain, eventually reaching the bottom. With ``Fast-IC" in  panel c we skip this slow evolution and convective plumes immediately cross the entire domain after convective onset and the dynamics quickly reach a converged state. We show probablity density functions (PDFs) of vertical velocity, $w$, for both cases in panel e. The two PDFs are distinct; the adiabatic-IC case has a narrower distribution and smaller average velocities, where the PDF of the Fast-IC case has a wider distribution which is close to the final velocity distribution seen in panel f. In the right column we show dynamics for the adiabatic-IC case (panel b) and the fast-IC case (panel d) as well as PDFs for both cases (panel f) at $t=6000$. At this time the dynamics and structure have converged for both initial condition choices. The dynamics not only look similar between panels b and d, but the distribution of velocities have converged in panel f. The ``Fast-IC'' case reached a final distribution with a mean of -0.013, standard deviation of 0.048, skewness of -4.24, and kurtosis of 21.74. The adiabatic initial conditions reached a final state with a mean of -0.012, standard deviation of 0.048, skewness of -4.10, and kurtosis of 20.37.

\newpage 

\section{Table of Simulations}
\label{app:big_table}
In table \ref{tab:all_sims} we display all simulations used in this study where we show control parameters $\mathrm{Ra}/\mathrm{Ra}_\mathrm{crit}$ and $\mathcal{H}$, horizontal resolution $n_\mathrm{horiz}$ and vertical resolution $n_z$, dimensionality, heating timescale $t_heat$, simulation run time $t_\mathrm{sim}$, averaging window $t_\mathrm{avg}$, and time and volume averaged Reynolds number, Mach number, Nusslet Number, and change in entropy across the domain. All timescales are expressed in nondimensionalized time units.
\begin{table}
    \centering
    
\begin{longtable} {c|c|c|c|c|c|c|c|c|c|c|c}
$\mathrm{Ra}/\mathrm{Ra}_\mathrm{crit}$ & $\mathcal{H}$ & $n_\mathrm{horiz}$ & $n_\mathrm{z}$ & 2D/3D & $t_\mathrm{heat}$ & $t_\mathrm{sim}$ & $t_\mathrm{avg}$ & Re & Ma & Nu & $\Delta s$ \\
\hline
$1.94 \times 10^{2}$ & ~0.029 &  128 &  64 &  2D & 13.465 & 120791 & 13465 & $2.08 \times 10^{1}$ & 0.019 & $2.17 \times 10^{0}$ & -0.043 \\
$1.95 \times 10^{2}$ & ~0.167 &  128 &  64 &  2D & 5.657 & ~51409 & ~5656 & $2.08 \times 10^{1}$ & 0.043 & $2.22 \times 10^{0}$ & -0.235 \\
$1.95 \times 10^{2}$ & ~0.400 &  128 &  64 &  2D & 3.651 & ~33683 & ~3651 & $2.07 \times 10^{1}$ & 0.065 & $2.29 \times 10^{0}$ & -0.531 \\
$1.93 \times 10^{2}$ & ~0.750 &  128 &  64 &  2D & 2.667 & ~24755 & ~2666 & $2.01 \times 10^{1}$ & 0.083 & $2.35 \times 10^{0}$ & -0.929 \\
$1.85 \times 10^{2}$ & ~1.333 &  128 &  64 &  2D & 2.000 & ~19339 & ~2000 & $1.94 \times 10^{1}$ & 0.100 & $2.46 \times 10^{0}$ & -1.481 \\
$1.96 \times 10^{2}$ & ~2.500 &  128 &  64 &  2D & 1.506 & ~14235 & ~1506 & $2.05 \times 10^{1}$ & 0.120 & $2.71 \times 10^{0}$ & -2.291 \\
$1.93 \times 10^{2}$ & ~6.000 &  128 &  64 &  2D & 1.040 & ~10680 & ~1040 & $2.20 \times 10^{1}$ & 0.138 & $3.24 \times 10^{0}$ & -3.780 \\
$1.91 \times 10^{2}$ & 10.667 &  128 &  64 &  2D & 0.839 & ~~8444 & ~~839 & $3.39 \times 10^{1}$ & 0.143 & $4.18 \times 10^{0}$ & -4.631 \\
$1.94 \times 10^{3}$ & ~0.029 &  256 &  128 &  2D & 19.764 & 130839 & 19763 & $5.87 \times 10^{1}$ & 0.015 & $3.25 \times 10^{0}$ & -0.031 \\
$1.95 \times 10^{3}$ & ~0.167 &  256 &  128 &  2D & 8.303 & ~51280 & ~8302 & $5.95 \times 10^{1}$ & 0.036 & $3.33 \times 10^{0}$ & -0.169 \\
$1.95 \times 10^{3}$ & ~0.400 &  256 &  128 &  2D & 5.359 & ~33869 & ~5359 & $5.88 \times 10^{1}$ & 0.055 & $3.43 \times 10^{0}$ & -0.385 \\
$1.93 \times 10^{3}$ & ~0.750 &  256 &  128 &  2D & 3.914 & ~24627 & ~3914 & $5.95 \times 10^{1}$ & 0.075 & $3.55 \times 10^{0}$ & -0.673 \\
$1.85 \times 10^{3}$ & ~1.333 &  256 &  128 &  2D & 2.935 & ~18330 & ~2935 & $5.84 \times 10^{1}$ & 0.094 & $3.75 \times 10^{0}$ & -1.081 \\
$1.96 \times 10^{3}$ & ~2.500 &  256 &  128 &  2D & 2.210 & ~13516 & ~2210 & $6.29 \times 10^{1}$ & 0.120 & $4.16 \times 10^{0}$ & -1.696 \\
$1.93 \times 10^{3}$ & ~6.000 &  256 &  128 &  2D & 1.526 & ~10251 & ~1526 & $6.93 \times 10^{1}$ & 0.146 & $5.04 \times 10^{0}$ & -2.879 \\
$1.91 \times 10^{3}$ & 10.667 &  256 &  128 &  2D & 1.232 & ~~2464 & ~1232 & $8.05 \times 10^{1}$ & 0.164 & $6.00 \times 10^{0}$ & -3.783 \\
$1.93 \times 10^{3}$ & 16.500 &  256 &  128 &  2D & 1.064 & ~~7618 & ~1064 & $9.73 \times 10^{1}$ & 0.177 & $7.12 \times 10^{0}$ & -4.453 \\
$1.94 \times 10^{4}$ & ~0.029 &  256 &  128 &  2D & 29.010 & 143950 & 29008 & $1.88 \times 10^{2}$ & 0.014 & $5.25 \times 10^{0}$ & -0.020 \\
$1.95 \times 10^{4}$ & ~0.167 &  256 &  128 &  2D & 12.187 & ~59992 & 12187 & $1.87 \times 10^{2}$ & 0.034 & $5.33 \times 10^{0}$ & -0.112 \\
$1.95 \times 10^{4}$ & ~0.400 &  256 &  128 &  2D & 7.866 & ~38995 & ~7866 & $1.87 \times 10^{2}$ & 0.051 & $5.43 \times 10^{0}$ & -0.260 \\
$1.93 \times 10^{4}$ & ~0.750 &  256 &  128 &  2D & 5.745 & ~29062 & ~5744 & $1.87 \times 10^{2}$ & 0.069 & $5.58 \times 10^{0}$ & -0.463 \\
$1.85 \times 10^{4}$ & ~1.333 &  256 &  128 &  2D & 4.309 & ~22096 & ~4309 & $1.87 \times 10^{2}$ & 0.090 & $5.80 \times 10^{0}$ & -0.761 \\
$1.96 \times 10^{4}$ & ~2.500 &  256 &  128 &  2D & 3.244 & ~16738 & ~3244 & $2.00 \times 10^{2}$ & 0.115 & $6.43 \times 10^{0}$ & -1.219 \\
$1.93 \times 10^{4}$ & ~6.000 &  256 &  128 &  2D & 2.240 & ~11800 & ~2240 & $2.36 \times 10^{2}$ & 0.160 & $7.78 \times 10^{0}$ & -2.144 \\
$1.91 \times 10^{4}$ & 10.667 &  256 &  128 &  2D & 1.809 & ~~9610 & ~1809 & $2.80 \times 10^{2}$ & 0.189 & $9.26 \times 10^{0}$ & -2.887 \\
$1.93 \times 10^{4}$ & 16.500 &  256 &  128 &  2D & 1.562 & ~~8306 & ~1562 & $3.28 \times 10^{2}$ & 0.210 & $1.07 \times 10^{1}$ & -3.509 \\
$1.94 \times 10^{5}$ & ~0.029 &  512 &  256 &  2D & 42.581 & ~68091 & 15773 & $5.85 \times 10^{2}$ & 0.014 & $8.23 \times 10^{0}$ & -0.013 \\
$1.95 \times 10^{5}$ & ~0.167 &  512 &  256 &  2D & 17.888 & ~29625 & ~6722 & $5.80 \times 10^{2}$ & 0.032 & $8.33 \times 10^{0}$ & -0.075 \\
$1.95 \times 10^{5}$ & ~0.400 &  512 &  256 &  2D & 11.546 & ~19045 & ~4323 & $5.89 \times 10^{2}$ & 0.049 & $8.46 \times 10^{0}$ & -0.176 \\
$1.93 \times 10^{5}$ & ~0.750 &  512 &  256 &  2D & 8.432 & ~13706 & ~3165 & $5.82 \times 10^{2}$ & 0.066 & $8.66 \times 10^{0}$ & -0.317 \\
$1.85 \times 10^{5}$ & ~1.333 &  512 &  256 &  2D & 6.324 & ~10594 & ~2433 & $5.93 \times 10^{2}$ & 0.089 & $8.93 \times 10^{0}$ & -0.532 \\
$1.96 \times 10^{5}$ & ~2.500 &  512 &  256 &  2D & 4.761 & ~~7847 & ~1856 & $6.36 \times 10^{2}$ & 0.116 & $9.70 \times 10^{0}$ & -0.877 \\
$1.93 \times 10^{5}$ & ~6.000 &  512 &  256 &  2D & 3.288 & ~~5614 & ~1307 & $7.61 \times 10^{2}$ & 0.167 & $1.17 \times 10^{1}$ & -1.596 \\
$1.91 \times 10^{5}$ & 10.667 &  512 &  256 &  2D & 2.655 & ~~4103 & ~1078 & $9.02 \times 10^{2}$ & 0.203 & $1.37 \times 10^{1}$ & -2.211 \\
$1.93 \times 10^{5}$ & 16.500 &  512 &  256 &  2D & 2.293 & ~~3960 & ~~929 & $1.06 \times 10^{3}$ & 0.229 & $1.60 \times 10^{1}$ & -2.721 \\
$1.94 \times 10^{6}$ & ~0.029 &  1024 &  512 &  2D & 62.500 & ~19419 & ~3870 & $1.72 \times 10^{3}$ & 0.012 & $1.33 \times 10^{1}$ & -0.009 \\
$1.95 \times 10^{6}$ & ~0.167 &  1024 &  512 &  2D & 26.255 & ~~8315 & ~1633 & $1.71 \times 10^{3}$ & 0.029 & $1.29 \times 10^{1}$ & -0.048 \\
$1.95 \times 10^{6}$ & ~0.400 &  1024 &  512 &  2D & 16.948 & ~~5513 & ~1075 & $1.73 \times 10^{3}$ & 0.045 & $1.32 \times 10^{1}$ & -0.114 \\
$1.93 \times 10^{6}$ & ~0.750 &  1024 &  512 &  2D & 12.377 & ~~4030 & ~~798 & $1.74 \times 10^{3}$ & 0.061 & $1.37 \times 10^{1}$ & -0.208 \\
$1.85 \times 10^{6}$ & ~1.333 &  1024 &  512 &  2D & 9.283 & ~~3129 & ~~604 & $1.73 \times 10^{3}$ & 0.081 & $1.39 \times 10^{1}$ & -0.356 \\
$1.96 \times 10^{6}$ & ~2.500 &  1024 &  512 &  2D & 6.988 & ~~2361 & ~~460 & $1.91 \times 10^{3}$ & 0.110 & $1.55 \times 10^{1}$ & -0.600 \\
$1.93 \times 10^{6}$ & ~6.000 &  1024 &  512 &  2D & 4.826 & ~~1669 & ~~329 & $2.20 \times 10^{3}$ & 0.156 & $1.81 \times 10^{1}$ & -1.140 \\
$1.91 \times 10^{6}$ & 10.667 &  1024 &  512 &  2D & 3.896 & ~~1364 & ~~261 & $2.63 \times 10^{3}$ & 0.194 & $2.04 \times 10^{1}$ & -1.607 \\
$1.93 \times 10^{6}$ & 16.500 &  1024 &  512 &  2D & 3.365 & ~~1167 & ~~230 & $3.12 \times 10^{3}$ & 0.222 & $2.51 \times 10^{1}$ & -2.016 \\
$1.94 \times 10^{7}$ & ~0.029 &  2048 &  1024 &  2D & 91.737 & ~10311 & ~7033 & $4.81 \times 10^{3}$ & 0.011 & $2.29 \times 10^{1}$ & -0.006 \\
$1.95 \times 10^{7}$ & ~0.167 &  2048 &  1024 &  2D & 38.537 & ~~7507 & ~2888 & $4.83 \times 10^{3}$ & 0.026 & $2.22 \times 10^{1}$ & -0.031 \\
$1.95 \times 10^{7}$ & ~0.400 &  2048 &  1024 &  2D & 24.876 & ~~4682 & ~1895 & $4.90 \times 10^{3}$ & 0.040 & $2.56 \times 10^{1}$ & -0.075 \\
$1.93 \times 10^{7}$ & ~0.750 &  2048 &  1024 &  2D & 18.167 & ~~2443 & ~1395 & $4.89 \times 10^{3}$ & 0.055 & $2.21 \times 10^{1}$ & -0.138 \\
$1.85 \times 10^{7}$ & ~1.333 &  2048 &  1024 &  2D & 13.625 & ~~1905 & ~1072 & $4.81 \times 10^{3}$ & 0.073 & $1.96 \times 10^{1}$ & -0.237 \\
$1.96 \times 10^{7}$ & ~2.500 &  2048 &  1024 &  2D & 10.257 & ~~1223 & ~~795 & $5.04 \times 10^{3}$ & 0.094 & $2.52 \times 10^{1}$ & -0.405 \\
$1.93 \times 10^{7}$ & ~6.000 &  2048 &  1024 &  2D & 7.084 & ~~1527 & ~~568 & $5.86 \times 10^{3}$ & 0.133 & $2.47 \times 10^{1}$ & -0.771 \\
$1.91 \times 10^{7}$ & 10.667 &  2048 &  1024 &  2D & 5.719 & ~~1135 & ~~435 & $7.06 \times 10^{3}$ & 0.170 & $4.57 \times 10^{1}$ & -1.116 \\
$1.93 \times 10^{7}$ & 16.500 &  2048 &  1024 &  2D & 4.939 & ~~~675 & ~~358 & $8.36 \times 10^{3}$ & 0.204 & $5.23 \times 10^{1}$ & -1.436 \\
\hline
$1.93 \times 10^{2}$ & ~0.750 &  128 &  64 &  3D & 2.667 & ~~2667 & ~1332 & $1.94 \times 10^{1}$ & 0.081 & $2.72 \times 10^{0}$ & -0.835 \\
$1.93 \times 10^{3}$ & ~0.750 &  256 &  128 &  3D & 3.914 & ~~3538 & ~1767 & $4.63 \times 10^{1}$ & 0.064 & $4.18 \times 10^{0}$ & -0.596 \\
$1.94 \times 10^{4}$ & ~0.029 &  256 &  128 &  3D & 29.010 & ~~5541 & ~2756 & $1.06 \times 10^{2}$ & 0.010 & $6.30 \times 10^{0}$ & -0.018 \\
$1.93 \times 10^{4}$ & ~0.750 &  256 &  128 &  3D & 5.745 & ~~1172 & ~~583 & $1.07 \times 10^{2}$ & 0.049 & $6.67 \times 10^{0}$ & -0.416 \\
$1.91 \times 10^{4}$ & 10.667 &  256 &  128 &  3D & 1.809 & ~~~904 & ~~112 & $1.62 \times 10^{2}$ & 0.130 & $1.11 \times 10^{1}$ & -2.559 \\
$1.93 \times 10^{5}$ & ~0.750 &  512 &  256 &  3D & 8.432 & ~~2340 & ~~219 & $2.59 \times 10^{2}$ & 0.040 & $1.05 \times 10^{1}$ & -0.269
\end{longtable}
    \caption{Table of all simulations used in this paper.}
    \label{tab:all_sims}
\end{table}

\newpage

\ULforem
\clearpage
%

\end{document}